# ARTICLE

# Observation of an anisotropic Dirac cone reshaping and ferrimagnetic spin polarization in an organic conductor

Michihiro Hirata[1,2,†], Kyohei Ishikawa[1], Kazuya Miyagawa[1], Masafumi Tamura[3], Claude Berthier[2], Denis Basko[4], Akito Kobayashi[5], Genki Matsuno[5] & Kazushi Kanoda[1]

The Coulomb interaction among massless Dirac fermions in graphene is unscreened around the isotropic Dirac points, causing a logarithmic velocity renormalization and a cone reshaping. In less symmetric Dirac materials possessing anisotropic cones with tilted axes, the Coulomb interaction can provide still more exotic phenomena, which have not been experimentally unveiled yet. Here, using site-selective nuclear magnetic resonance, we find a non-uniform cone reshaping accompanied by a bandwidth reduction and an emergent ferrimagnetism in tilted Dirac cones that appear on the verge of charge ordering in an organic compound. Our theoretical analyses based on the renormalization-group approach and the Hubbard model show that these observations are the direct consequences of the long-range and short-range parts of the Coulomb interaction, respectively. The cone reshaping and the bandwidth renormalization, as well as the magnetic behaviour revealed here, can be ubiquitous and vital for many Dirac materials.

[1] Department of Applied Physics, University of Tokyo, Bunkyo-ku, Tokyo 113-8656, Japan. [2] Laboratoire National des Champs Magnétiques Intenses, UPR 3228 CNRS, EMFL, UGA, UPS and INSA, Boite Postale 166, Grenoble, Cedex 9 38042, France. [3] Department of Physics, Faculty of Science and Technology, Tokyo University of Science, Noda, Chiba 278-8510, Japan. [4] Université Grenoble Alpes and CNRS, Laboratoire de Physique et Modélisation des Milieux Condensés UMR 5493, 25 rue des Martyrs, Grenoble 38042, France. [5] Department of Physics, Nagoya University, Chikusa-ku, Nagoya 464-8602, Japan. † Present address: Institute for Materials Research, Tohoku University, Aoba-ku, Sendai 980-8577, Japan. Correspondence and requests for materials should be addressed to M.H. (email: michihiro_hirata@imr.tohoku.ac.jp) or to K.K. (email: kanoda@ap.t.u-tokyo.ac.jp).

Dirac materials[1] are a novel class of solid-state systems, in which the low-energy electronic excitations are described by pseudo-relativistic massless Dirac fermions (DFs). Triggered by the studies in two-dimensional (2D) graphene[2] and the surface of three-dimensional topological insulators[3], extended now to three-dimensional Weyl and Dirac semimetals with strong spin–orbit coupling[4–6], many intriguing properties of DFs have been revealed and have constituted active topics in modern condensed matter physics. The role of Coulomb interaction is one of such issues of particular interest[7–17]. For instance, in charge-neutral 2D massless DF systems, composed of two gapless points at $E_F$, the long-range (LR) part of the Coulomb potential $V(\mathbf{q})$ ($\mathbf{q}$: wave vector) is unscreened owing to the vanishing density of states (DOS) at $E_F$. Consequently, the LR character of the interaction ($V(\mathbf{q}) \propto 1/|\mathbf{q}|$) is preserved at low energy, which couples to the fermionic excitations and induces a logarithmic correction to the Fermi velocity $v_F$ and associated physical quantities[7–9]. The logarithmic velocity renormalization induces a nonlinear reshaping of the cones around each of Dirac points (DPs), as observed in graphene near the charge neutrality point[11–14].

However, graphene is a special case of 2D massless DF systems, in which isotropic Dirac cones with vertical axes have the DPs at particularly symmetric points on the Brillouin zone boundary[2]. Indeed, theoretical studies have revealed that massless DFs possessing anisotropic cones and the DPs at arbitrary $\mathbf{k}$-points emerge more generally in a broad class of materials[1,18–20]. A typical example in 2D is the organic-layered compound α-(BEDT-TTF)$_2$I$_3$ (α-I$_3$) (BEDT-TTF = bis(ethylenedithio)-tetrathiafulvalene) (Fig. 1a), which has a pair of Dirac cones occurring at two distinct points ($\pm \mathbf{k}_D$) in the 2D first Brillouin zone (Fig. 1c)[20–30]. The electronic structure of α-I$_3$, described on the base of molecular orbitals as usual in this type of compounds[31], is rather involved compared with graphene due to the presence of four sites per unit cell (Fig. 1b) with anisotropic hopping amplitudes[32,33]. The system has only the inversion symmetry[32–34], which, in conjunction with the anisotropic hopping, brings about a tilt of Dirac cones and drives the 2D DPs away from high crystallographic symmetry positions[20,21,23,25,35,36] (Fig. 1c). A remarkable feature is that, because of the 3/4-filled nature of the electronic bands[23–26,33], the two gapless points are anchored at $E_F$ by this band filling in α-I$_3$.

Another issue of great physical interest in α-I$_3$ in terms of the Coulomb interaction phenomena is that, within the pressure–temperature ($P$–$T$) phase diagram (Fig. 1g)[32,37–41], the 2D massless DF phase appears in the vicinity of an insulating phase with charge order, as first pointed out by transport measurements[22]. This contrasts with the case of graphene, in which no phase transitions have been observed at least in the absence of a quantizing magnetic field[10]. The charge-ordered phase in α-I$_3$, which is induced by the strong short-range (SR) electron correlations in this 3/4-filled system[40,41], is suppressed when applying a $P$ above a critical value of $P_C \approx 1.2$ GPa (Fig. 1g) and turns into the 2D massless DF phase[22,39]. Once the high-$P$ phase is reached, the Dirac cones become stable against further pressurization; in fact, the gapless point is fixed at $E_F$ on varying the hopping integrals in a finite range by virtue of the 3/4-filled nature of the electronic bands, as revealed by band-structure calculations[23,25,26,33,42–45]. The presence of such a phase transition in this system potentially offers the possibility to test the impact of the SR electron correlations on the behaviours of 2D massless DFs. Moreover, the tilt of anisotropic Dirac cones[36] coupled with the SR and LR parts of the Coulomb interaction opens new possibilities in the physics of 2D massless DFs. For instance, it is predicted to bring about a non-uniform reshaping of titled cones[46], novel non-Fermi liquid behaviours near the quantum critical point[16,17], where two DPs merge[47,48], enhanced shot noise for quantum transport[49] and anomalous charge/spin textures inside the unit cell[48,50]. Studying the electronic structures and the role of the Coulomb interaction in pressurized α-I$_3$, which remains unclear up to date, is thus of primary importance to understand the various effects of the Coulomb interaction in 2D massless DFs.

In this article, we focus on the 2D massless DF phase in α-I$_3$ emerging under a hydrostatic pressure ($P > P_C$) and present experimental evidence for interaction effects of massless DFs. Employing site-selective nuclear magnetic resonance (NMR), we uncover three distinct interaction phenomena induced by the electron–electron Coulomb interaction. First, NMR-shift measurements in conjunction with renormalization-group (RG) analyses reveal a $T$-driven cone reshaping around each of the DPs due to the LR part of the Coulomb interaction. Because of this reshaping, tilted cones become effectively isotropic at low energies. Second, quantitative RG analyses establish that the best fit to the data inevitably requires a strong bandwidth reduction inherent to the SR electron correlation, as often discussed in strongly correlated materials. Finally, an anomalous ferrimagnetic spin polarization is observed, which is accounted for by the onsite Coulomb repulsion, as revealed by a simulation based on the Hubbard model presented here. These experimental and theoretical investigations demonstrate that α-I$_3$ under $P$ is an intrinsically interacting 2D massless DF system, in which both the LR and SR parts of the Coulomb interaction strongly influence the electronic behaviours.

## Results

**Basic principles to probe tilted Dirac cones.** Our strategy to investigate tilted Dirac cones in α-I$_3$ is as follows. The crystal structure of α-I$_3$ has a 2D unit cell with four molecular sites (dubbed sites $A$, $A'$ ($=A$), $B$ and $C$), each of which constitutes a sublattice in the crystalline $ab$-plane (Fig. 1a,b). The four molecular orbitals on these sites form a pair of tilted Dirac cones near $E_F$ (Fig. 1c). Around the gapless point at $E_F$, a very unique situation is realized where the Bloch state has different weights in the amplitudes of the four molecular orbitals. The band-structure calculation[25] revealed that these weights, dubbed site-spectral weights hereafter, show anisotropic $\mathbf{k}$ dependence around each of 2D DPs with a clear contrast between non-equivalent sites. The corresponding site-spectral weight for the sublattice $j = A$ ($=A'$), $B$ and $C$ around the DP at $\mathbf{k}_D$, $n_j^\zeta(\mathbf{q})$ (equation (11)), is shown in Fig. 1d–f, where $\mathbf{q} = (q_x, q_y)$ is defined as $\mathbf{q} = \mathbf{k} - \mathbf{k}_D$ and $\zeta = \pm$ is the band index (Fig. 1c). (For details, see Methods.) Notably, the anisotropy of the site-spectral weights makes a particular distinction between the site $B$ and $C$. Namely, the Bloch electrons with large $v_F$ (in the steep slope of the tilted cones) have a large weight on the $B$-site wavefunction, whereas the Bloch states with small $v_F$ (on the opposite side of the cones in the gentle slope) have a large weight on the $C$-site wavefunction (Fig. 1e,f and Supplementary Fig. 1a,c). Thus, if one probes the local electronic states on sites $B$ and $C$ separately by means of a site-selective measurement, it is possible to reveal the electronic nature of the Bloch states in the steep part and the gentle part of the tilted Dirac cones individually. Taking advantage of this feature, we performed a site-selective NMR in this compound to separately elucidate the electronic states in the two slopes of the Dirac cones. Specifically, the Knight shift, derived from the NMR line shift measured at a temperature $T$, is converted into the local electron-spin susceptibility on the site $j$, $\chi_s^j(T)$, which is given by a thermal average of the site-spectral weight $n_j^\zeta(\mathbf{q})$ around $E_F$ summed over all $\mathbf{q}$ for both electrons ($\zeta = +$) and holes ($\zeta = -$). Hence, the site-selective NMR in α-I$_3$ works as an effectively

**q**-resolved probe of 2D DFs thermally excited around each of DPs. Indeed, the electronic excitations in the steep and gentle parts of the Dirac cones can be almost independently probed by $\chi_s^B$ and $\chi_s^C$, respectively, as we will demonstrate below.

**NMR observation of tilted Dirac cones.** To address the electron interaction issues of 2D massless DFs, we have carried out

$^{13}$C-NMR measurements in $\alpha$-I$_3$ at $P = 2.3$ GPa ($> P_{\rm C}$; see Fig. 1g) on the four molecular sites in the unit cell, $j = A$, $A'$ ($= A$), $B$ and $C$. The two $^{13}$C nuclei (spin $I = 1/2$) introduced at the centre of BEDT-TTF (ET) molecules (inset of Fig. 1a) are used for $^{13}$C NMR, which are known as a sensitive probe of electronic states at $E_F$ in this class of compounds[51]. Figure 2a shows the typical NMR spectra observed at a magnetic field of

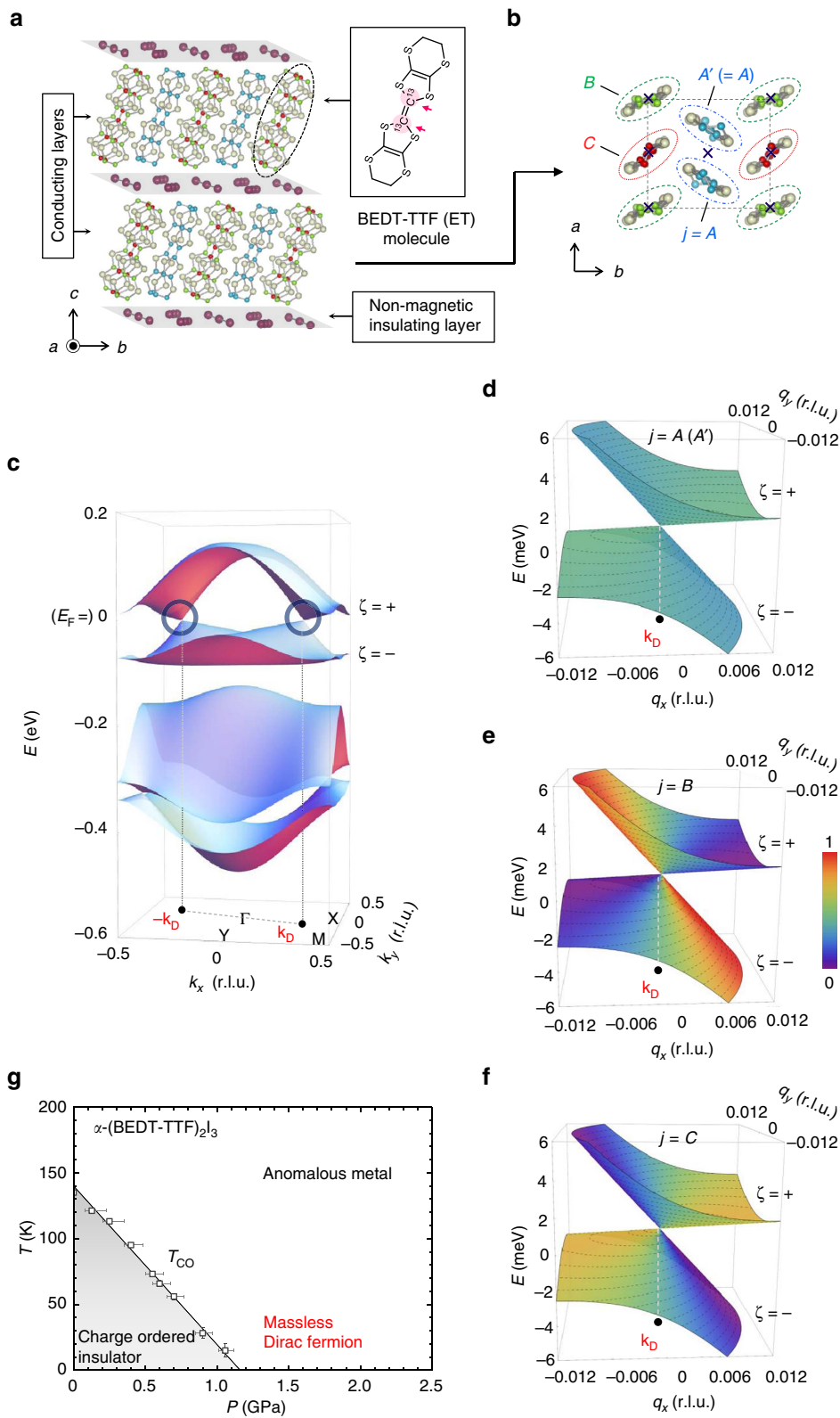

$H = 6\,T$, applied in the crystalline $ab$-plane. Eight lines are observed in the spectra, associated with the four molecular sites in the unit cell. By changing the field orientation in the crystalline $ab$-plane and examining the angular dependence of the line positions, the eight lines are to be assigned to two doublets from sites $B$ and $C$ and one quartet from the site $A$ ($= A'$)[37]. (The site $A$ and $A'$ are not distinguished hereafter). Figure 2b shows the typical angular dependence of the NMR total shift for each molecular site, defined as the centre-of-mass position of the doublet or the quartet. The $j$-site total shift for a given $T$ and a field-angle $\psi$ (measured from the crystalline $a$ axis; Fig. 2c) is expressed as $S_j(T, \psi) = \bar{A}_j(\psi)\chi_s^j(T) + \sigma_j(\psi)$, where the first term is the conduction-electron term (Knight shift) and the second term stands for the core-electron contribution (chemical shift). Note that the so-called hyperfine coupling constant, $\bar{A}_j(\psi)$, and $\sigma_j(\psi)$ are strongly $\psi$-dependent (with little $T$ and $P$ dependence), while

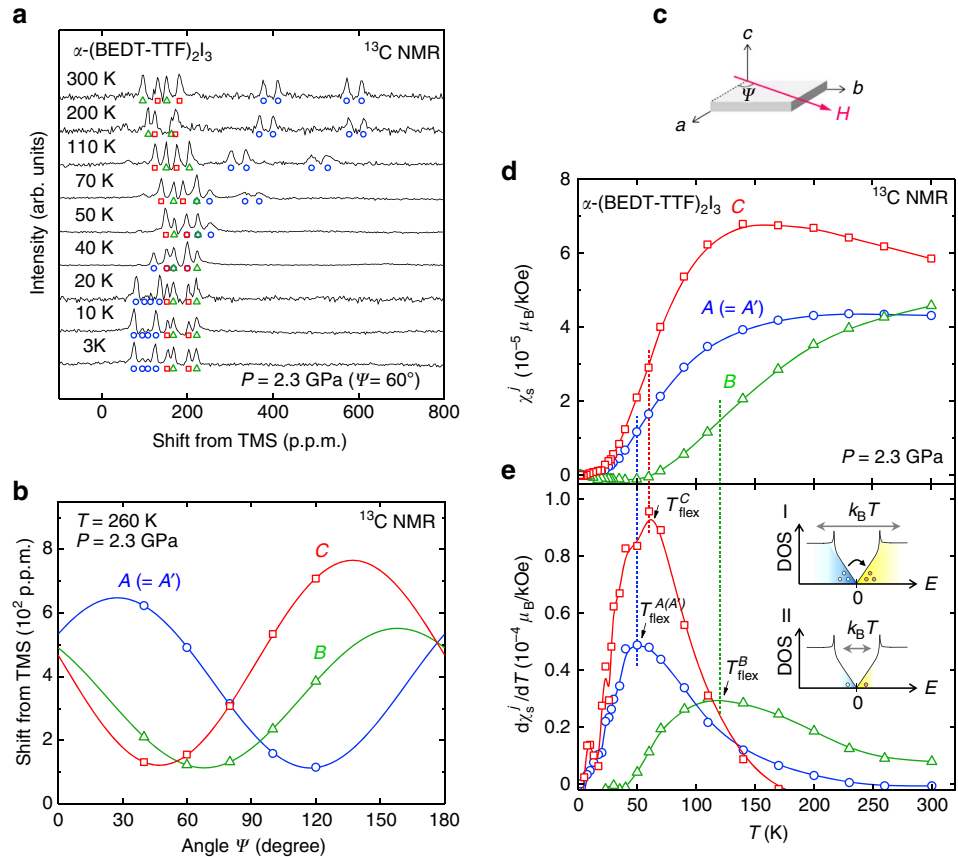

**Figure 2 | Site-selective NMR probes anisotropic tilted Dirac cones.** (**a**) Temperature, $T$, dependence of the $^{13}$C-NMR spectra measured under $P = 2.3\,GPa$ at the magnetic field of $H = 6\,T$ applied in the crystalline $ab$-plane (for the direction $\psi = 60°$). Symbols denote the $^{13}$C lines associated to the three non-equivalent sites in the unit cell: $j = A$ ($= A'$) (circles), $B$ (triangles) and $C$ (squares). (**b**) Field angular dependence of the centre-of-mass positions of the corresponding $^{13}$C lines for each sublattice. Lines indicate the sinusoidal fitting curves. (**c**) Definition of the field-angle $\psi$ in the $ab$-plane, measured from the $a$ axis. (**d**) Local electron-spin susceptibility $\chi_s^j(T)$ plotted against $T$, which are determined from the spectra measured at $\psi = 60°$ for the site $A$ ($= A'$) and $\psi = 120°$ for the site $B$ and $C$. Lines are the guide to the eyes. (**e**) The $T$ dependence of the first derivative $d\chi_s^j/dT$. The inflection point, $T_{flex}^j$, ($T_{flex}^A \approx 50\,K$, $T_{flex}^B \approx 120\,K$ and $T_{flex}^C \approx 60\,K$) is indicated by arrows and vertical dashed lines. Lines stand for the guides to the eyes. Inset: Schematic illustrations of thermal excitations of electron-hole pairs around the gapless point at $E_F$ ($= 0$) for high $T$ ($T > T_{flex}^j$) (I) and low $T$ ($T < T_{flex}^j$) (II).

---

**Figure 1 | Tilted Dirac cones in $\alpha$-(BEDT-TTF)$_2$I$_3$.** (**a**) Layered structure of $\alpha$-(BEDT-TTF)$_2$I$_3$ ($\alpha$-I$_3$), with alternatingly stacked BEDT-TTF (ET) and triiodide (I$_3$) layers. Balls-and-stick diagram represents the molecular structures, where sticks indicate bonds; red, blue and green balls stand for carbon atoms; grey balls are sulphur atoms; and big purple balls indicate iodine atoms. One electron per two ETs is donated to a I$_3$ molecule due to charge transfer, constituting a quasi-2D (hole) conducting system in (ET)$_2^+$ layers and non-magnetic insulating I$_3^-$ layers. Inset of (**a**): The structure of a ET molecule. $^{13}$C isotopes, substituted for $^{13}$C-NMR, are indicated by arrows. (**b**) 2D unit cell with four ET sites in the $ab$-plane, distinguished as the site $j = A$ ($= A'$) (blue), $B$ (green) and $C$ (red). Same colour correspondence as in **a**. The dashed rectangle indicates the unit cell, and crosses stand for the inversion centre. (**c**) Electronic band structure of $\alpha$-I$_3$ at high pressures ($P > P_C$; see **g**)[25]. As the band is 3/4-filled owing to the charge transfer, the Fermi energy $E_F$ ($= 0$) is present between the first ($\zeta = +$) and second ($\zeta = -$) bands from the top. Gapless points appear at $E_F$ and locate at $\pm \mathbf{k}_D$ in the first Brillouin zone, around which a pair of tilted Dirac cones are visible (indicated by circles). Wave vectors ($k_x$ and $k_y$) are given in the r.l.u. (**d–f**) The calculated titled Dirac cone around $E_F$ ($= 0$) plotted as a function of the wave vector $\mathbf{q} = \mathbf{k} - \mathbf{k}_D = (q_x, q_y)$. The label bar stands for the size of the site-spectral weight, $n_j^\zeta(\mathbf{q})$, for the non-equivalent site $j = A$ ($= A'$) (**d**), $B$ (**e**) and $C$ (**f**)[25] (equation (11) in Methods). The magnitude is normalized to the maximum value of $n_j^\zeta(\mathbf{q})$ for $j = B$. (**g**) The pressure–temperature ($P$–$T$) phase diagram. The charge ordering temperature $T_{CO}$ (squares) is determined from the electrical resistance ($R$) measurements[39], defined as the maximum of $-d(\ln R)/T$. Error bars follow those provided in ref. 39. At $T = 0$, the phase boundary locates at $P_C \approx 1.2\,GPa$ as determined from a linear extrapolation (solid line), corresponding to $dT_{CO}/dP \approx -110\,K\,GPa^{-1}$. r.l.u., reciprocal lattice unit.

the susceptibility $\chi_s^j(T)$ is isotropic in this compound[52]. Because the electronic excitations around the gapless point (at $E_F$) are vanishingly small at low temperatures, the total shift $S_j$ at the lowest $T$ ($=3$ K in the present experiment) is expected to provide the chemical shift term $\sigma_j$. Thus, subtracting $\sigma_j(\psi)$ from $S_j(T, \psi)$ and employing the value of the hyperfine coupling constant $A_j(\psi)$ reported at ambient pressure[37], the total shift $S_j(T, \psi)$ is converted into the local electron-spin susceptibility $\chi_s^j(T)$ (for details, see Methods).

Figure 2d shows the temperature dependence of $\chi_s^j(T)$ at the site $j = A$, $B$ and $C$, which arises from the inter-band thermal (electron–hole) excitations across the gapless point at $E_F$ and is proportional to the $k_B T$ average of the $j$-site DOS around $E_F$: $\chi_s^j(T) \propto \langle D_j(E) \rangle_{k_B T}$. Decreasing from $T = 300$ K, $\chi_s^j(T)$ exhibits $T$-independent features down to $T \approx 200$ K, followed by a rapid decrease with a clear difference in the size of the susceptibility between non-equivalent sites, $\chi_s^C > \chi_s^A > \chi_s^B$ (see also Supplementary Fig. 2), and finally becomes vanishingly small at all sites. The observation, in particular the crossover from $\chi_s^j \sim$ const. to $\chi_s^j \sim 0$ on cooling, indicates that there is an energy-independent large $D_j(E)$ at high energies (inset I of Fig. 2e) and a vanishingly small $D_j(E)$ at low energies around $E_F$ (inset II of Fig. 2e). This is consistent with the band-structure calculations[24,25,50], where a flat DOS is predicted at high energies above the van-Hove singularity (locating at about 12 meV off $E_F$) and linear energy dependence is suggested around the band-crossing point at $E_F$. Note that $D_j(E)$ around the gapless point ($E = E_F = 0$) is given by a $\mathbf{q}$ summation of the site-spectral weight $n_j^\zeta(\mathbf{q})$ (Fig. 1d–f) at a given energy $E$ for the band $\zeta$ (equation (14)). Then, the anisotropic $\mathbf{q}$-dependence of $n_j^\zeta(\mathbf{q})$ around the DPs of tilted Dirac cones (Supplementary Fig. 1) is expected to bring about $D_C(E) > D_A(E) > D_B(E)$ at low energy. This is in excellent agreement with the observed relation $\chi_s^C > \chi_s^A > \chi_s^B$ below $T \approx 200$ K (Fig. 2d) and is the direct consequence of the fact that the site $B$ and $C$ selectively probes the steep and gentle slopes of titled cones, respectively, consistent with the prediction of the effective tight-binding (TB) model given in ref. 25. All these observations demonstrate the existence of tilted Dirac cones with $E_F$ located at the gapless point.

**Fermi velocity renormalization.** However, the strong nonlinear temperature dependence in $\chi_s^j(T)$ below $T \approx 150$ K does not comply with the expectation of TB calculations, which leads to $\chi_s^j \propto T$ at low temperatures[25,50]. To better visualize this point, we plot the first derivative of the susceptibility $(d\chi_s^j/dT)$, as shown in Fig. 2e. With decreasing $T$, $d\chi_s^j/dT$ exhibits a peak at $T_{flex}^B \approx 120$ K for the site $B$, and at $T_{flex}^{A,C} \approx 50$–60 K for the site $A$ and $C$, and then drops continuously to zero at all sites towards low temperatures. These features are in striking contrast to the TB calculation[25], where $d\chi_s^j/dT$ increases on cooling but saturates at low $T$ (Supplementary Fig. 3d–f). Indeed, $\chi_s^j(T)$ varies almost quadratic in $T$ at all sites below the inflection point ($T_{flex}^j$), which suggests a nonlinear suppression of $D_j(E)$ around $E_F$ ($=0$) in an energy range of $\Delta E^j \approx k_B T_{flex}^j$. As the total DOS, $D(E)$, is proportional to the inverse square of $v_F$ in 2D massless DF systems for the non-interacting case $D(E) = |E|/\pi\hbar^2 v_F^2$ (refs 2,20), a suppression of DOS in turn corresponds to an enhancement of $v_F$. Thus, the observed peak structure in $d\chi_s^j/dT$ strongly indicates that a $T$-driven renormalization of $v_F$ grows below $T \approx T_{flex}^j$. The most probable origin of this effect is the LR part of the Coulomb interaction between electrons, which is unscreened at $E_F$ in charge-neutral massless DF systems and is known to cause a logarithmic correction to $v_F$ either driven by tuning carrier densities[1,7–13] or temperatures[9,46]. We recall, however, the value of $T_{flex}^j$ is twice higher for the site $B$ ($T_{flex}^B \approx 120$ K) than for the site $A$ and $C$ ($T_{flex}^{A,C} \approx 50$–60 K). At first glance, this may add an extra complication to the data interpretation but in fact turns out to be a direct consequence of the anisotropy of the site-spectral weight $n_j^\zeta(\mathbf{q})$ (Fig. 1d–f) and the tilt of Dirac cones, as we shall see below.

**Renormalization-group analyses.** To further understand the nonlinear temperature dependence of $\chi_s^j(T)$ at each site, we have examined the self-energy correction effect due to the LR Coulomb interaction. For this, we employed a RG approach based on an effective Hamiltonian near the gapless point[20,25], whose energy-momentum dispersion is given by

$$E_\pm(\mathbf{q}) = \hbar \left( \mathbf{w}_0 \cdot \mathbf{q} \pm \sqrt{v_x^2 q_x^2 + v_y^2 q_y^2} \right), \qquad (1)$$

where $\mathbf{w}_0 = (w_{0x}, w_{0y})$ and $\mathbf{v} = (v_x, v_y)$ are velocities reflecting the tilt and anisotropy of the cone, respectively (for details, see Methods). At the one-loop level, the self-energy correction leads to a renormalization of $\mathbf{v}$ but does not affect $\mathbf{w}_0$ (Supplementary Fig. 4a)[46]. The RG flow of $\mathbf{v}$ is expressed as

$$
\begin{aligned}
\frac{1}{v_x}\frac{dv_x}{dl} &= \frac{8}{\pi^2 N}\int_0^{2\pi}\frac{d\varphi}{2\pi}2\cos^2\varphi F(g_\varphi), \\
\frac{1}{v_y}\frac{dv_y}{dl} &= \frac{8}{\pi^2 N}\int_0^{2\pi}\frac{d\varphi}{2\pi}2\sin^2\varphi F(g_\varphi),
\end{aligned}
\qquad (2)
$$

where $N = 4$ is the number of fermion species, corresponding to two DPs and two spin projections, $\mathbf{q} = q(\cos\varphi, \sin\varphi)$ is measured from $\mathbf{k}_D$, $l = \ln(\Lambda/q)$ is the momentum scale, $\Lambda$ ($= 0.667\,\text{Å}^{-1}$) is a momentum cutoff of the size of the inverse lattice constant[33] and is circular around the DP, $g_\varphi = 2\pi e^2 N/(16\,\varepsilon\sqrt{v_x^2\sin^2\varphi + v_y^2\cos^2\varphi})$ is the coupling, $F(g_\varphi)$ has the form $F(g_\varphi) = (-\pi/2 + g_\varphi + \arccos g_\varphi/\sqrt{1 - g_\varphi^2})/g_\varphi$ and $\varepsilon$ is the dielectric constant. (Note that equation (2) is obtained in the leading order in $1/N$ assuming $N \gg 1$, which is valid both for the weak and strong Coulomb interaction.)

Assuming the four velocities given by the effective TB calculation ($\mathbf{w}_0^{TB}$ and $\mathbf{v}^{TB}$)[25] as initial velocities at $q = \Lambda$, we have calculated the RG correction effects on $\chi_s^j(T)$ (Fig. 3a–c) $D_j(E)$ (Fig. 4a–c) and the energy spectrum (Fig. 4d–f). (For the justifications of employing this TB model as well as the velocities $\mathbf{w}_0^{TB}$ and $\mathbf{v}^{TB}$ in performing RG calculations, see Methods.) Here, we note that the temperature is used as an explicit scale parameter in the calculation of $\chi_s^j(T)$ that determines the RG flow. To get a reasonable agreement between the calculation and experiment, a phenomenological parameter $u$ is introduced to adjust the velocities at $q = \Lambda$ such that $\mathbf{w}_0' = u\mathbf{w}_0^{TB}$ and $\mathbf{v}' = u\mathbf{v}^{TB}$. The two parameters in the calculation, $(u, \varepsilon)$, are then optimized from a least-square fit to the experimental susceptibilities. Good agreements are obtained in the fit especially at the site $A$ and $C$ (Fig. 3a–c), which lead $(u, \varepsilon) \approx (0.35, 1)$. (Note that the fitting results are sensitive to the choice of $u$ while they are little dependent on $\varepsilon$ in the range $\varepsilon \approx 1$–30; for details, see Methods and Supplementary Figs 4–7). The calculation demonstrates that the nonlinear $T$ dependence of $\chi_s^j(T)$ below $T \approx T_{flex}^j$ can be properly ascribed to the logarithmic renormalization of $v_F$. In Fig. 4a–c, the calculated shape of $D_j(E)$ is shown around the gapless point at $E_F$. A strong suppression from the $E$-linear DOS is seen at low energies due to the renormalization. Figure 4d–f, depicts the corresponding energy spectrum around the DP (at $\mathbf{k}_D$), where the colours indicate the magnitude of the site-spectral weight, $n_j^\zeta(\mathbf{q})$, in Fig. 1d–f, respectively. A nonlinear reshaping of the tilted cone induced by the renormalization is clearly visible around the gapless point. It should be stressed that a good agreement is accomplished only when a small value of $u$ ($\approx 0.35$) is used. The fact that we have $u < 1$ indicates a reduction of the initial velocities or of the hopping amplitudes between the adjacent lattice sites. In conventional strongly correlated materials, the SR part of the Coulomb interaction is well known to induce this sort of hopping (or bandwidth) renormalization due to the frequency dependence of the self-energy[53]. We believe that the observed $u$-reduction effect in the initial velocities occurs because of this self-energy correction due to the SR Coulomb interaction, which is not considered implicitly in the original RG calculation.

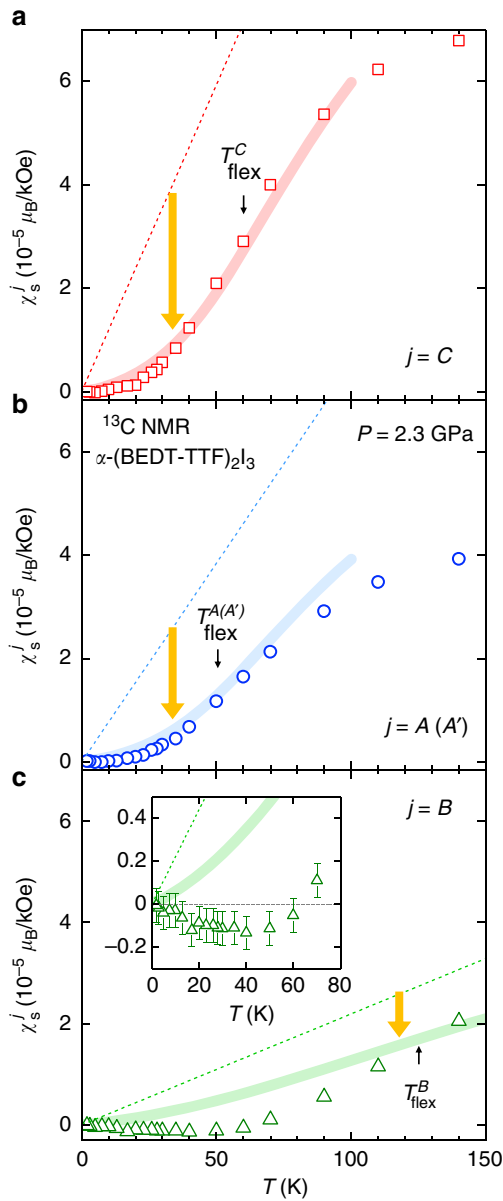

**Figure 3 | Temperature driven velocity renormalization.** (a–c) Temperature dependence of the calculated spin susceptibility in the RG approach (bold curves) for the non-equivalent site $j = C$ (a), $A$ (=$A'$) (b) and $B$ (c) in the unit cell, plotted together with the experimental $\chi_s^j(T)$ (symbols). Thin arrows indicate the inflection point, $T_{flex}^j$. As the non-interacting reference in the RG theory, we assumed the effective TB model of ref. 25. The susceptibilities are calculated as discussed in the text by employing the momentum cutoff $\Lambda = 0.667\,\text{Å}^{-1}$ and the optimum parameters ($u$, $\varepsilon$) = (0.35, 1), determined from the fit to the data (Methods). The dotted lines ($\chi_s^j \propto T$) indicates the expected $T$ dependence for gapless excitations around the DP in the absence of the RG correction[68], where the bandwidth reduction effect is incorporated (through the parameter $u$). Thick bold arrows represent the suppression of $\chi_s^j(T)$ with respect to $\chi_s^j \propto T$ due to the RG flow of the velocities, where the velocities flow towards larger $\ln(\Lambda/|\mathbf{q}|)$ (longer wavelengths) on decreasing $T$. That is, $T$ plays the role of the flow parameter in the experiment, and the RG flow manifests itself typically in the energy range of $\Delta E^j \approx k_B T_{flex}^j$ around $E_F$. The insets of c: the low-$T$ close-up of $\chi_s^j$. Error bars stand for the s.e.m.

Remarkably, the calculation well reproduces the observed difference in the thermal energy scale of the $T$-driven renormalization in $\chi_s^j(T)$, $\Delta E^j$ ($\approx k_B T_{flex}^j$), at the site $B$ and ($A$, $C$) as indicated by thin arrows in Fig. 3a–c. This distinction stems from the tilt of the Dirac cones and the resultant momentum dependence of the energy cutoff $\delta$ around the DP, $\delta(\varphi, \zeta) = \hbar v_F(\varphi, \zeta)\Lambda$, where $v_F(\varphi, \zeta)$ is the $\varphi$ dependent Fermi velocity in the band $\zeta$ (Supplementary Fig. 1a,b). Namely, the energy cutoff is large for the large-$v_F$ DFs, dominantly probed by the site $B$, while it is small for the small-$v_F$ DFs, having a large weight on the site $C$ (and $A$). Hence, the renormalization starts from a higher $T$ in $\chi_s^B(T)$ than in $\chi_s^C(T)$ (and $\chi_s^A(T)$), producing the observed energy scale difference $\Delta E^B > E^{A,C}$, consistent with the previous RG calculation of Isobe et al.[46]. It is also worth mentioning that tilted cones become more isotropic at lower energies near $E_F$ because of the non-uniform velocity renormalization around each DP, as reflected in Fig. 4d–f. This is because the anisotropic term in the Hamiltonian is small ($v_x/v_y \approx 1$) in $\alpha$-I$_3$ (ref. 25), and we have $|\mathbf{w}_0| \ll |\mathbf{v}|$ around the DP due to the RG flow (Supplementary Fig. 4a), leading the tilting term ($\mathbf{w}_0$) to be effectively negligible near $E_F$.

From all these, it is concluded that our RG analyses appropriately capture many of the essential parts of the experimental results. They constitute experimental evidence for the bandwidth renormalization (the $u$-reduction effect) due to the SR repulsion between electrons as well as the $T$-driven logarithmic renormalization of $v_F$ by the LR part of the Coulomb interaction. Nevertheless, we note that, at low temperatures, the agreement is less satisfactory for the site $B$ compared with the other sites (Fig. 3c), suggesting the presence of another correlation effect. Indeed, we will clarify this point by a lattice-model simulation, as described below.

**Emergent ferromagnetic spin polarization.** The temperature dependence of $\chi_s^B(T)$ in the experiment is appreciably stronger and more complex than what is predicted by the RG calculation (Fig. 3c). Indeed, the experimental $\chi_s^B(T)$ exhibits an anomalous sign change at $T \approx 60\,\text{K}$ and an upturn with a negative slope below $T \approx 40\,\text{K}$ (inset of Fig. 3c), while the RG calculation shows monotonic temperature dependence. The observation of $\chi_s^B < 0$ is in sharp contrast to $\chi_s^A > 0$ and $\chi_s^C > 0$ in the experiment (Fig. 5a), indicating an emergent ferrimagnetic spin polarization in which the local magnetic field points antiparallel to the applied field at the site $B$ while it is parallel to the field at all other sites (Fig. 6). To further understand this sublattice-scale magnetism, we have investigated the Hubbard model with an onsite repulsive (Hubbard) interaction, $U$, at a mean-field level within the random phase approximation (RPA). For the RPA calculation, we have considered both the inter-band and intra-band contributions to the spin susceptibility with a wave vector $\mathbf{Q} = \mathbf{0}$ (for details, see Methods). Figure 5b presents the calculated temperature dependence of the RPA spin susceptibility at the site $B$. Using $U = 0.14\,\text{eV}$, the RPA calculation (in particular the inter-band term) clearly reproduces the observed negative spin susceptibility for site $B$ ($\chi_s^B < 0$) at similar temperatures. Moreover, the negative susceptibility appears only at site $B$ in the calculation (Supplementary Fig. 3a–c) in good agreement with the experiment (Fig. 5a and Supplementary Fig. 2). These facts show that the ferrimagnetic polarization is induced by the onsite Hubbard interaction. The fact that the negative polarization emerges solely on the site $B$ might be relevant to a superexchange-like interaction between the site $A$ and $A'$ (via $B$) (see Fig. 6). Density functional calculation[24] suggests a largest hopping amplitude on this path (b2 in Supplementary Fig. 8), and X-ray and Raman scattering measurements[32,38] point to the largest hole density (despite the small spin density) at the intermediate site $B$ in the unit cell. Then, if there is an antiferromagnetic (ferrimagnetic) coupling between sites $A$ and $B$ ($A'$ and $B$), a large energy gain is expected due to the kinetic energy of electrons, which favours the observed pattern of the ferrimagnetic polarization.

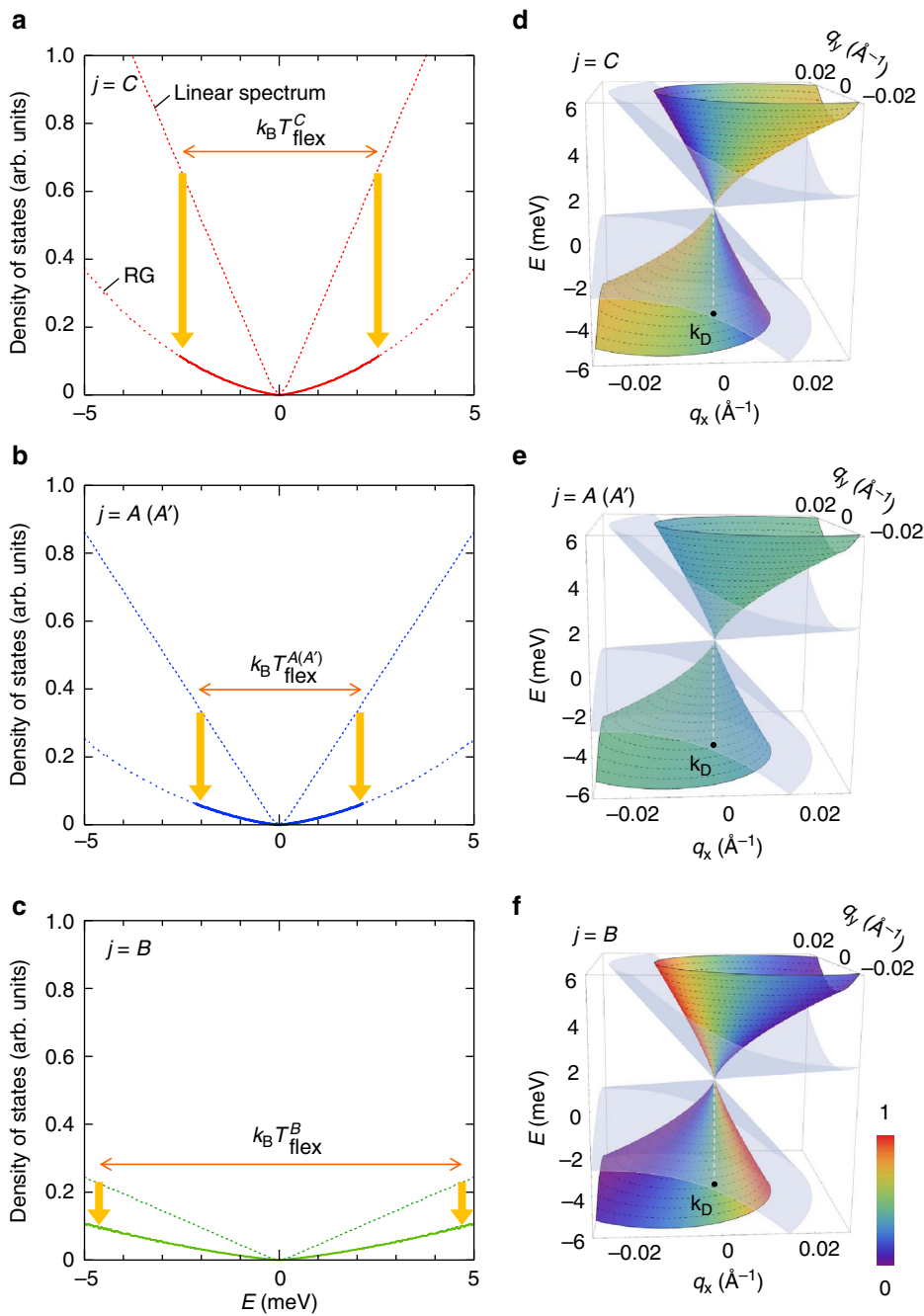

**Figure 4 | LR Coulomb-induced tilted Dirac cone reshaping. (a–c)** The calculated DOS near the gapless point (at $E = E_F = 0$) plotted as a function of energy $E$ for the non-equivalent site $j = C$ **(a)**, $A$ ($= A'$) **(b)** and $B$ **(c)** in the unit cell. The thin dotted and thick dashed curves are the DOS profiles for the linear spectrum and the RG-corrected DOS, respectively, calculated for the momentum cutoff $\Lambda = 0.667\,\text{Å}^{-1}$ and optimum fitting parameters $(u, \varepsilon) = (0.35, 1)$. Note that the bandwidth reduction effect, associated with $u = 0.35$ ($<1$), is taken into account in both curves. The thick bold parts in the RG-corrected DOS, highlighted around the gapless point, correspond to the energy range ($\Delta E^j \approx k_B T_{\text{flex}}^j$) where the $T$-driven RG flow is visible in the experiment (also indicated by the horizontal left-right arrows). Vertical thick bold arrows represent the suppression of the DOS due to the LR Coulomb interaction. **(d–f)** Calculations of the reshaped tilted Dirac cone (around $\mathbf{k}_D$) induced by the RG flow of the velocities, derived from the same parameters as in **a–c**. The label bar reflects the site-spectral weight $\eta_j^i(\mathbf{q})$ around the cone for the non-equivalent site $j = C$ **(d)**, $A$ ($= A'$) **(e)** and $B$ **(f)** in the unit cell plotted as a function of the wave vector $\mathbf{q} = \mathbf{k} - \mathbf{k}_D = (q_x, q_y)$ (in $\text{Å}^{-1}$). The outer grey cone stands for the linear spectrum in the absence of the LR Coulomb interaction, where the bandwidth reduction effect is considered (corresponding to the $E$-linear DOS in **a–c**). (See Methods for details.)

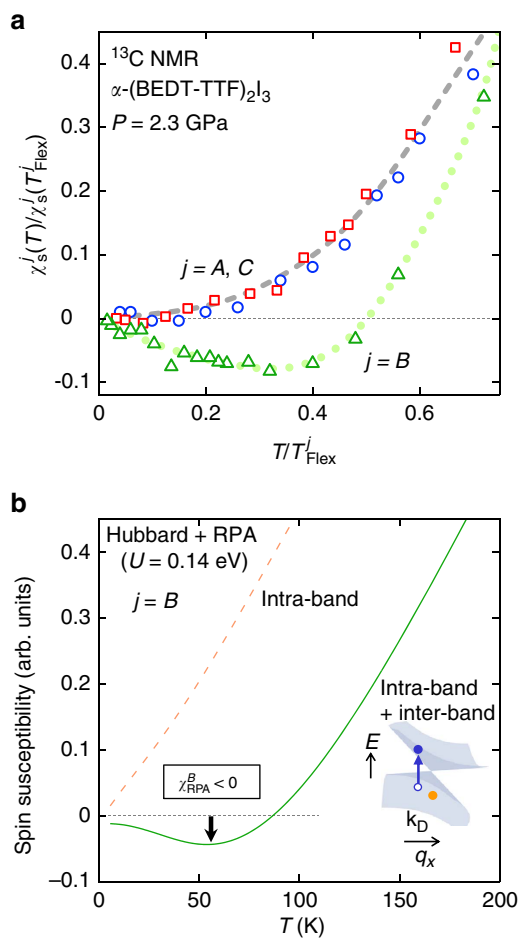

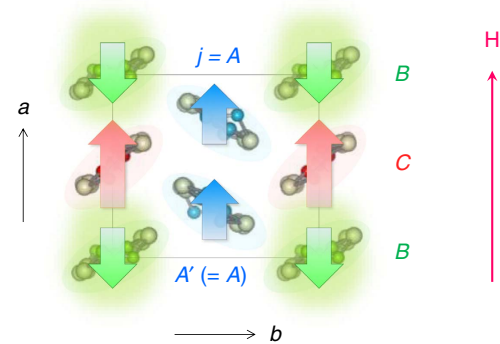

**Figure 6 | Ferrimagnetic spin polarization.** Schematic illustration of the ferrimagnetic spin polarization suggested at low temperatures (below 60 K) by our site-selective susceptibility measurements. Thick arrows represent the direction of the local magnetic field on the non-equivalent site $A (= A')$, $B$ and $C$ in the unit cell, which is opposite to the external field direction ($\mathbf{H} \| a$ in this figure) at the site $B$ while it is parallel at all other sites.

**Figure 5 | Emergent negative spin susceptibility at the site $B$.**
(**a**) Observed local electron spin susceptibility $\chi_s^j(T)$ normalized to the value at the inflection point $T_{\text{flex}}^j$, $\chi_s^j(T)/\chi_s^j(T_{\text{flex}}^j)$, plotted against the normalized temperature, $T/T_{\text{flex}}^j$, for the non-equivalent site $j = A(A')$ (circles), $B$ (triangles) and $C$ (squares) in the unit cell. Dotted and dashed curves are guide to the eyes. (**b**) Calculated $T$ dependence of the $B$-site-spin susceptibility $\chi_s^B$ in the RPA based on the Hubbard model. The total susceptibility (solid curve) and the intra-band contribution (dashed curve) for the onsite Hubbard interaction of $U = 0.14$ eV are shown. Inset: Intra/inter-band thermal excitations with the wave vector $\mathbf{Q} = \mathbf{0}$ in the Dirac cone considered for the RPA calculation. The Fermi energy $E_F$ locates at the band-crossing gapless point

Generally speaking, RPA tends to overestimate the effect of correlations, as it does not consider the self-energy correction to the energy bands[54]. In fact, the onsite interaction we use, $U = 0.14$ eV, is chosen smaller than what would be typically used ($U \approx 0.4$ eV) (refs 25,26). This discrepancy brings another support that the self-energy correction due to the SR interaction is of great importance in this system, consistent with the bandwidth reduction effect we discussed in the RG calculation. Finally, we note that RPA is unable to reproduce the observed nonlinear $T$ dependence of $\chi_s^j(T)$ at low temperatures. This is because the self-energy correction due to the LR Coulomb interaction, which is the main origin of the nonlinear $\chi_s^j(T)$, is not taken into account in RPA (for details, see Methods and Supplementary Fig. 3d–f).

## Discussion
So far, we have demonstrated three distinct Coulomb-interaction phenomena in the 2D massless DF phase of $\alpha$-$I_3$, which develop systematically at different temperature scales (or energy scales of thermal excitations), as summarised in Table 1. A bandwidth renormalization (or the $u$-reduction of $v_F$) occurs due to the SR Coulomb interaction which appears to exist from room temperature down to lowest $T$. At temperatures $T \lesssim T_{\text{flex}}^j$ (or in the corresponding energy range around the gapless point at $E_F$), a $T$-driven logarithmic renormalization of $v_F$ and the resultant non-uniform reshaping of the Dirac cones appear, due to the LR part of the Coulomb interaction. With further decreasing $T$, a ferrimagnetic spin polarization shows up in the unit cell because of the onsite Coulomb repulsion between electrons.

First, we mention that the observed non-uniform reshaping of titled Dirac cones in $\alpha$-$I_3$ should affect other physical observables at low temperature or at low magnetic field. As the cones become effectively isotropic around each of DPs (Fig. 4d–f), the shape of the cross-section of the cone is different at high energy and close to the gapless point, which should cause a $T$-dependent anisotropy of the in-plane electrical conductivity. Another experiment that would be able to see the reshaping is infrared spectroscopic measurements, known as a powerful tool to reveal the Landau level (LL) structure in graphene[55]. In a perpendicular magnetic field ($H_\perp$) normal to the 2D plane, the massless DFs in $\alpha$-$I_3$ exhibit the LL spectrum, $E_{\zeta,n} = \zeta \hbar l_B^{-1} \sqrt{v_x v_y} (1 - \Omega^2)^{3/4} \sqrt{2|n|}$ (refs 56,57), where $\zeta = \pm$ distinguishes the electron and hole bands crossing at the gapless point, $n$ ($= 0, \pm 1, \pm 2, \cdots$) is the LL index, $l_B = (\hbar/eH_\perp)^{1/2}$ is the magnetic length and $\Omega = \sqrt{(w_{0x}/v_x)^2 + (w_{0y}/v_y)^2}$ is a tilting parameter. As $E_F$ locates at the gapless point in $\alpha$-$I_3$ and the $n = 0$ LL is half-filled, one may be able to detect, for instance, the dipole transition $n = 0 \to n = 1$ in the absorption line at the energy $\Delta E_{10} = \hbar l_B^{-1} \sqrt{2v_x v_y} (1 - \Omega^2)^{3/4}$. The velocities ($v_x$, $v_y$) with a logarithmic correction may appear in $\Delta E_{10}$ at low field; $v_x$ ($\approx v_y$) will increase by a factor of four by changing $\ln(\Lambda/q)$ from 1 to 5 (Supplementary Fig. 4a) and $(1 - \Omega^2)^{3/4}$ increases by a factor of two (from $\sim 0.5$ to $\sim 1.0$). One would expect to see this change in $\Delta E_{10}$ at low temperatures where the LL broadening, mainly caused by the thermal scattering of carriers in the $n = 0$ LL (refs 27,28), becomes sufficiently small.

Theoretically, the renormalization of the coupling constant of the Coulomb interaction (the RG flow of $v_F$) makes the system flow to a weak coupling regime at low energies (or at low $T$ in the case of a $T$-driven RG flow as in our experiment)[10]. As organic compounds are very clean and are little influenced by impurities, this consequence of the RG flow implies that low-$T$ electron

**Table 1 | Interaction effects of Dirac fermions in α-(BEDT-TTF)₂I₃.**

| T (K) | Long-range Coulomb | Short-range Coulomb | Emergent interaction effects |
|---|---|---|---|
| $\Delta E/k_B \leq 300$ | — | $\omega$-dependence of self-energy | Bandwidth renormalization ($u$-reduction) |
| $\Delta E/k_B \leq T^j_{\rm Seg}$ ($\approx$60–120) | Self-energy effect due to $V(\mathbf{q}) \propto 1/|\mathbf{q}|$ | — | Logarithmic correction to $v_F$ |
| $\Delta E/k_B \leq 60$ | — | Onsite Hubbard $U$ | Ferrimagnetic spin polarization ($\chi^s_S < 0$) |

BEDT-TTF, bis(ethylenedithio)-tetrathiafulvalene.
Here, $\Delta E$ ($=k_B T$) indicates the energy range around the gapless point at $E_F$, within which the inter-band thermal excitations are effective (see the inset of Fig. 2e). $T^j_{\rm Seg}$ stands for the peak temperature in the first derivative of the electron spin susceptibility $\chi^j_s$ (Fig. 2e) at the non-equivalent site $j = A$ ($= A'$), $B$ and $C$ in the unit cell (Fig. 1b), $\omega$ indicates the frequency and $u$ represents the phenomenological parameter introduced in the RG analyses (see the text).

correlation effects, if any, would be induced by the SR part of the Coulomb interaction, as usually the case in conventional strongly correlated materials. In fact, the insulating phase possessing charge order[22,26,31–41] emerges next to the 2D massless DF phase on the $P$–$T$ phase diagram in α-I₃ (Fig. 1g), suggesting the vital role of the SR Coulomb interactions in the massless DF phase of this 3/4-filled system, in line with recent mean-field calculations[48]. In the half-filled system counterpart, strong electron correlations are predicted to stabilize Mott insulators and charge density waves[10]. However, the typical experimental conditions seem to locate away from these situations[1,10,58], and no phase transition has been reported yet. Under a strong magnetic field, on the other hand, a gapped liquid phase is observed with a quantized Hall conductivity at fractional filling factors, stabilized by the SR part of the Coulomb interaction[10]. Similar physics may occur in thin films of α-I₃ as well, where the integer quantum Hall effects have recently been observed[59].

To conclude, our NMR measurements combined with theoretical calculations have demonstrated three $T$-dependent Coulomb interaction effects of 2D massless DFs (Table 1) in pressurized α-I₃ ($P > P_C$), having tilted Dirac cones and gapless points fixed at $E_F$. We found that the LR part of the Coulomb interaction, which is unscreened around the gapless DPs, causes a $T$-driven renormalization of the velocity and induces a non-uniform reshaping of tilted Dirac cones. Quantitative analyses of the cone reshaping based on the RG approach further necessitates a large bandwidth reduction due to the SR electron correlation. Moreover, we showed that the onsite Coulomb repulsion gives rise to a ferrimagnetic spin polarization as unveiled by the numerical calculations using the Hubbard model. These findings can be distinguished from the case of weakly interacting 2D massless DFs in graphene with vertical Dirac cones and are consistent with the emergent correlated phase on the verge of the massless DF phase in the $P$–$T$ phase diagram. Continuing this study to the vicinity of the phase transition at $P_C$ ($\approx$1.2 GPa) would be of particular interest, which may connect the physics of the massless DFs and conventional strongly correlated materials.

## Methods

**Sample preparation.** Single crystals of α-(BEDT-TTF)₂I₃ (α-I₃) (ref. 60) with the dimensions of $0.1 \times 0.5 \times 2.0\,{\rm mm}^3$ were synthesized from $^{13}$C-enriched BEDT-TTF (ET) molecules using the conventional electrochemical method. To perform $^{13}$C-NMR measurements, the central carbon atoms of BEDT-TTF molecules connected by a double bond were 99% enriched by carbon-13 ($^{13}$C) isotopes (inset of Fig. 1a) with a nuclear spin $I = 1/2$.

**Pressurization scheme.** A hydrostatic pressure of $P = 2.3$ GPa was applied to the sample using a BeCu/NiCrAl clamp-type pressure cell, with the Daphne 7373 oil as a pressure medium. At this pressure, the oil locates close to the liquid–solid phase transition point at room temperature[61,62]. To avoid applying uniaxial strains to the sample, the cell was kept at a sufficiently high temperature ($T \approx 50\,^{\circ}$C) during the pressurization. With decreasing temperature from $T = 300$–3 K, a pressure reduction of $\Delta P \sim 0.1$ GPa occurs inside the cell[61,62]. The inner pressure at the lowest $T$ we measured (3 K) is, however, substantially higher than the transition pressure to the charge-ordered phase at $T \approx 0$ ($P_C \approx 1.2$ GPa; see Fig. 1g)[22,30,39–41,63,64], indicating that the $P$-reduction will not change the physics. Hence, we neglect this effect throughout the paper.

**$^{13}$C-NMR measurements.** $^{13}$C-NMR measurements were performed in α-I₃ in a magnetic field $H$ of 6 T applied parallel to the crystalline $ab$-plane (Fig. 1a). To get NMR signals, the standard spin-echo techniques were used with a commercially available homodyne spectrometer. Spin-echo signals were recorded at a fixed radio frequency after the conventional spin-echo pulse sequence of $t_{\pi/2} - \tau - t_\pi$ (with $\tau = 5$–25 μs and $t_{\pi/2} - t_\pi = 0.6$–1.2 μs) and were converted into the NMR spectra via Fourier transformation. The resonance frequency of the natural abundance $^{13}$C nuclei in TMS (tetramethylsilane (CH₃)₄Si) was used as the origin of the NMR shift. We note that the present work is targeted at lower $T$ and higher $P$ compared with the earlier NMR studies[65,66], which is more suitable for exploring the nature the low-$T$ 2D massless DF phase emerging at $P > P_C$ on the phase diagram (Fig. 1g)[22,27–30,39].

**Line assignments of the NMR spectra.** The details of line assignments for the $^{13}$C-NMR spectra are given elsewhere[37,65]. Here, we shall describe only the essence. The $^{13}$C-NMR spectra of α-I₃ show large temperature dependence (Fig. 2a and Supplementary Fig. 9) and field-angle ($\psi$) dependence (Supplementary Fig. 10). They have eight $^{13}$C lines that consist of two doublets (from the $B$ and the $C$ molecules) and one quartet (from the $A$ ($= A'$) molecule) (Fig. 2a). The doublet and the quartet are caused by the nuclear dipole–dipole interaction in a ET molecule (between the two $^{13}$C nuclear magnetic moments around the molecular centre; see the inset of Fig. 1a)[37,51,65]. The NMR total shift $S_j$ for a particular site $j$ is determined from the centre-of-mass position of the corresponding $^{13}$C lines, which is expressed by a sum of the Knight shift $K_j (= \bar{A}_j \chi^j_s)$ and the chemical shift $\sigma_j$ terms, $S_j(T, \psi) = K_j(T, \psi) + \sigma_j(\psi)$, as mentioned in the text. (The orbital van-Vleck contribution is negligible in ET-based salts because of the low lattice symmetry[51].) Both $K_j(T, \psi)$ and $\sigma_j(\psi)$ are highly anisotropic in this system[37,65–67], causing clear $\psi$-dependence of the spectra. The anisotropy of the shift $S_j(T, \psi)$ is, however, largely different between the molecule $A$ ($A'$) and molecules $B$ and $C$, reflecting the fishbone-like arrangement of ET molecules in the crystalline $ab$-plane (Fig. 1b). Thus, by rotating the magnetic field $H$ in this plane, one can assign the $^{13}$C lines to the different sublattices[37], with the aid of the X-ray diffraction data under pressure[33] (Supplementary Fig. 10).

**Conversion of the NMR shift to the spin susceptibility.** In α-I₃, the $T$-dependence curve of the $j$-site total shift $S_j(T, \psi)$ shows a largely distinct feature for different field orientations. For instance, the $T$ dependence of $S_{A(A')}(T, \psi)$ is large at $\psi = 60^\circ$ showing a prominent decrease with decreasing $T$ (Supplementary Fig. 11a), whereas it is small at $\psi = 120^\circ$ with an increase on cooling (Supplementary Fig. 11b). This difference can be accounted for by the anisotropy of the $T$-independent hyperfine coupling constant $\bar{A}_j(\psi)$, where $\bar{A}_j(\psi)$ can be either positive or negative depending on the value of $\psi$, as we will describe below. First, we note that the chemical shift $\sigma_j(\psi)$ is $T$-independent[51] as well as little affected by $P$ (see Supplementary Discussion). Thus, the observed $T$ dependence of the total shift $S_j$ can be ascribed to the $T$-variation of the Knight shift $K_j = \bar{A}_j \chi^j_s$. Here, $\bar{A}_j$ is the hyperfine-coupling constant averaged for the two central $^{13}$C nuclei in the molecule $j$ (Fig. 1a), which is $\psi$-dependent reflecting the anisotropy of the coupling tensor[33,37,66,67]. The principal values of the tensor are, however, weakly affected by the change of $T$ and $P$ (Supplementary Figs 12 and 13). Moreover, the spin susceptibility $\chi^j_s$ is expected to be isotropic in this compound[52]. These points clearly indicate that $K_j$ can practically be expressed as $K_j(T, \psi) = \bar{A}_j(\psi) \chi^j_s(T)$ (for details, see Supplementary Discussion).

Notice that there are no excitations around the gapless DP in the ground state of massless DF systems. This means, at the lowest $T$, the Knight shift $K_j(T, \psi)$ is expected to become vanishingly small[68], and the total shift $S_j(T, \psi)$ resumes to the $T$-independent chemical shift $\sigma_j(\psi)$ (Supplementary Fig. 12). For the site $B$, there is a negative slope in the $T$ dependence of the total shift $S_j(T, \psi)$ below $T \approx 40$ K, showing a small increase of the shift (of $\sim 5$ p.p.m.) towards lower $T$ (inset of Supplementary Fig. 11b). This is to be associated to the ferrimagnetic spin polarization, which causes a local magnetic field that points opposite to the external field only at the site $B$ (Fig. 6)[42]. The effect is, however, negligible at the lowest $T$ since the thermally excited polarization vanishes as $T \rightarrow 0$ (Fig. 5b and Supplementary Fig. 3c). Hence, we fitted the angular dependence of $S_j(T, \psi)$ for all sites at the lowest measured temperature ($T = 3$ K) and assumed this fitted curve to be the chemical shift value at each angle $\psi$. The total shift $S_j(T,\psi)$ for a particular site is then converted to $K_j(T,\psi)$ by subtracting this $\sigma_j(\psi)$. The subtraction is done at a field orientation where the total shift becomes close to the maximum in order to minimize the ambiguity of the chemical shift, namely, $\psi = 60°$ for the site $A(A')$ and $\psi = 120°$ for the site $B$ and $C$ (Supplementary Figs 11 and 12). The value of the hyperfine coupling constant $A_j(\psi)$ is calculated for these angles by employing the coupling tensors given at ambient pressure[37] by means of X-ray diffraction data reported at high pressure[33]. This yields $\bar{A}_{A(A')}(\psi = 60°) = 9.0$, $\bar{A}_B(\psi = 120°) = 6.4$ and $\bar{A}_C(\psi = 120°) = 9.9$ in kOe $\mu_B^{-1}$. In terms of these coupling constants, the Knight shift $K_j(T,\psi)$ is eventually converted to the local electron-spin susceptibility $\chi_s^j(T)$ (Fig. 2d).

**Effective tight-binding model for the tilted Dirac cone.** In order to construct reasonable arguments for the fitting analyses to the observed $\chi_s^j(T)$ (Fig. 2d), we have introduced a four-band band-structure calculation of ref. 25. The low-energy effective model based on this calculation shall be used as a non-interacting reference to the data analyses in particular in the RG calculation. (For details, see Methods: RG calculations). The rationales behind this choice are described here.

In $\alpha$-I$_3$, it has been shown that the gapless DPs at $\pm \mathbf{k}_D$ (Fig. 1c) are fixed at $E_F$ for a certain parameter region in the four-band TB parameter space with and without finite site potentials[23,33,42–45]. However, it is difficult to use bare TB parameters as adjustable variables in the fitting analyses of the low-$T$ part of $\chi_s^j(T)$ (in Fig. 2d). This is because the TB calculations lead to linear dispersion around the DP[21,23], which causes $\chi_s^j \propto T$, owing to the excitations around the gapless point at $E_F$ (ref. 68). Our experiment, on the other hand, exhibits nonlinear $T$-dependence in $\chi_s^j(T)$ at all sites below $T \approx T_{flex}^j$ (Fig. 2d,e) and a negative $\chi_s^B(T)$ at low $T$ ($\leq T_{flex}^B/2 \approx 60$ K; see the inset of Figs 3c and 5a). A model calculation based on simple linear dispersion can hardly account for these features. Thus, instead of fitting the data with TB parameters, we will use them as a minimal non-interacting reference and perform more sophisticated RG calculations based on a continuum model derived from that reference.

For the minimal model in our study, we use the band-structure calculation of ref. 25, which is practically based on a non-interacting TB model with adjustable site-dependent potentials, associated to the four molecular sites in the unit cell, $j = A(1)$, $A'(2)$, $B(3)$ and $C(4)$ (Fig. 1b). (This model shall be dubbed the effective TB model throughout this study.) Strictly speaking, this model takes into account the electron–electron interaction up to the nearest-neighbour terms and, at first glance, appears not to be a non-interacting model. However, as one works within a mean-field framework, the interaction merely ends up in additional site potentials in the expression of the Hamiltonian[40,41], thus making sense. In this sense, ref. 25 can be also considered as a TB model with adjustable site-dependent potentials from a practical point of view. Importantly, it is well known that the presence of the gapless point at $E_F$ is unaffected by modulations of this kind of site potentials within a range in this compound[19,33,42,43]. The chosen values of the site potentials in ref. 25, which are given by

$$I_1 = I_2 = 1.0964, \quad I_3 = 3.8755, \quad I_4 = 3.8277 \quad (\text{in eV}) \qquad (3)$$

are within this range and are thus acceptable. Using these potentials, the Hamiltonian of the effective TB model in ref. 25 can be eventually expressed by a $4 \times 4$ matrix, $\varepsilon_{ij}$ [$i,j = A(1)$, $A'(2)$, $B(3)$ and $C(4)$], whose Fourier transformed matrix elements are given by

$$\varepsilon_{ij}(\mathbf{k}) = t_{ij}(\mathbf{k}) + I_i \delta_{ij}, \qquad (4)$$

with the kinetic terms

$$
\begin{aligned}
t_{11}(\mathbf{k}) &= 2t_{a1'} \cos k_y, & t_{12}(\mathbf{k}) &= t_{a2} + \ = \ t_{a3} e^{-ik_y}, \\
t_{13}(\mathbf{k}) &= t_{b2} + t_{b3} e^{ik_x}, & t_{14}(\mathbf{k}) &= t_{b1} + t_{b4} e^{ik_x}, \\
t_{22}(\mathbf{k}) &= 2t_{a1'} \cos k_y, & t_{23}(\mathbf{k}) &= t_{b3} e^{ik_y} + t_{b2} e^{i(k_x + k_y)}, \\
t_{24}(\mathbf{k}) &= t_{b4} + t_{b1} e^{ik_y}, & t_{33}(\mathbf{k}) &= 2t_{a3'} \cos k_y, \\
t_{34}(\mathbf{k}) &= t_{a1} + t_{a1} e^{-ik_y}, & t_{44}(\mathbf{k}) &= 2t_{a4'} \cos k_y, \\
t_{ij}(\mathbf{k}) &= \left[ t_{ji}(\mathbf{k}) \right]^*, &
\end{aligned}
\qquad (5)
$$

The hopping integrals are better determined by *ab initio* calculations such that the resultant electronic bands become compatible with experimental observations in this system. For this, we employed the hopping integrals reported by the first-principle density-functional calculation at $T = 8$ K (ref. 24), as in ref. 25, which are given for the nearest neighbours by (in the unit of eV)

$$
\begin{aligned}
t_{a1}^{LT} &= -0.0267, \ t_{a2}^{LT} = -0.0511, \ t_{a3}^{LT} = 0.0323, \\
t_{b1}^{LT} &= 0.1241, \ t_{b2}^{LT} = 0.1296, \ t_{b3}^{LT} = 0.0513, \ t_{b4}^{LT} = 0.0152,
\end{aligned}
\qquad (6)
$$

and for the next nearest neighbours as

$$t_{a3'}^{LT} = 0.0119, \ t_{a4'}^{LT} = 0.0046, \ t_{a1'}^{LT} = 0.0060. \qquad (7)$$

(For the definition of the integrals, see Supplementary Fig. 8.) The largest integrals, $t_{b1}^{LT}$ and $t_{b2}^{LT}$, are known to vary about 15% by raising $T$ from 8 to 300 K (ref. 24), though the variation is less than a few per cent below 100 K. As our fitting analyses primarily focus on this low-$T$ region, it is reasonable to omit the $T$-dependence and keep using the hopping integrals estimated at $T = 8$ K, $\{t_p^{LT}; p = a1 - a4'\}$, at all $T$ as is done in ref. 25. By diagonalizing the Hamiltonian (equation (4)) in conjunction with the hopping integrals (equations (5)–(7)) and the site potentials (equation (2)), one obtains the four energy bands with tilted Dirac cones at $E_F$, as shown in Fig. 1c.

We note that it is very important to use these hopping integrals in equations (3) and (4) to reproduce the observed sign change of the Hall coefficient $R_H$ from $R_H > 0$ to $R_H < 0$ with decreasing $T^{69}$. (For details, see refs 26,70.)

From all these, we used the effective TB model in ref. 25 as our non-interacting reference to the RG analyses. It should be stressed that we do not mean to incorporate interaction effects at this level, and indeed the model we assumed is a purely non-interacting TB model with adjustable site-dependent potentials. Note that these values of site potentials are realistic because they lead to a site-dependent charge differentiation which is compatible with the observed X-ray and Raman scattering results in the conducting phase; see refs 25,32,34.

**Generalized Weyl Hamiltonian and site-spectral weight.** Around the band-crossing DPs, where the Fermi energy $E_F$ (put as $E = 0$ hereafter) is fixed in $\alpha$-I$_3$ due to the stoichiometry, the low-energy continuum model is shown to be given by the generalized Weyl Hamiltonian[20,23,25,35]

$$H = \hbar \left( \mathbf{w}_0 \cdot \mathbf{q} \sigma^0 + v_x q_x \sigma^x + v_y q_y \sigma^y \right), \qquad (8)$$

which describes the electronic states in the vicinity of one of the DPs. Here, $\mathbf{w}_0 = (w_{0x}, w_{0y})$ and $\mathbf{v} = (v_x, v_y)$ are effective velocities describing the tilt and the anisotropy of the Dirac cone, respectively; $\sigma^0$ is the $2 \times 2$ unit matrix; $(\sigma^x, \sigma^y)$ are the Pauli matrices; and $\mathbf{q} = (q_x, q_y)$ is the 2D wave vector measured from the DP at $\mathbf{k}_D$. Note that the twofold valley degeneracy associated to the two DPs at $\pm \mathbf{k}_D$ (Fig. 1c) will be omitted for simplicity, and we will hereafter focus on a single valley (at $\mathbf{k}_D$). (When the valley degeneracy is to be included in some of the expressions, we will specifically mention it.) The Hamiltonian (equation (8)) is defined in a space spanned by the Luttinger–Kohn bases[71]; $|\Phi_D^{LK1}\rangle$ and $|\Phi_D^{LK2}\rangle$. These bases are the two degenerate Bloch states at $\mathbf{k}_D$, which are given by a (normalized) superposition of the highest occupied molecular orbital $|h_j\rangle$ on each of the four different BEDT-TTFs (ETs) in the unit cell[36]

$$| \Phi_D^\lambda \rangle = \sum_{j=1}^4 \alpha_j^\lambda | h_j \rangle \qquad (\lambda = \text{LK1 or LK2}), \qquad (9)$$

where $j = A(1)$, $A'(2)$, $B(3)$ and $C(4)$ represents the different sites (see Fig. 1b). Diagonalization of equation (8) yields the eigenvalue $E_\zeta(\mathbf{q})$ (equation (1)) in terms of the two bands ($\zeta = \pm$) and the eigenstates (Goerbig, M.O., private communication.)

$$
\begin{aligned}
| \psi_\mathbf{q}^\zeta \rangle &= \frac{1}{\sqrt{2}} \left( \zeta | \Phi_D^{LK1} \rangle + e^{i\varphi_\mathbf{q}} | \Phi_D^{LK2} \rangle \right) \\
&= \frac{1}{\sqrt{2}} \sum_{j=1}^4 \left\{ \zeta \left( \alpha_j^{LK1} \right) + e^{i\varphi_\mathbf{q}} \left( \alpha_j^{LK2} \right) \right\} | h_j \rangle,
\end{aligned}
\qquad (10)
$$

where $\tan \varphi_\mathbf{q} = v_y q_y / v_x q_x$. We note that the two states $| \Phi_D^{LK1,LK2} \rangle$ in equation (10) have an equal weight for any value of $\mathbf{q}$, as in the two-band model of the graphene Dirac cone, whereas the four states $|h_{1,2,3,4}\rangle$ in equation 10 have not necessarily the same weight[25,72]. To see this, we define a (normalized) $\mathbf{q}$-dependent site-spectral weight by taking a projection of $|\psi_\mathbf{q}^\zeta\rangle$ onto $|h_j\rangle$, $n_j^\zeta(\mathbf{q}) = \left| \langle h_j | \psi_\mathbf{q}^\zeta \rangle \right|^2$ (ref. 36) (Goerbig, M.O., private communication.), which reads

$$
\begin{aligned}
n_j^\zeta(\mathbf{q}) &= \left| \frac{1}{\sqrt{2}} \left\{ \zeta \left( \alpha_j^{LK1} \right) + e^{i\varphi_\mathbf{q}} \left( \alpha_j^{LK2} \right) \right\} \right|^2 \\
&= \frac{1}{2} \left\{ \left| \alpha_j^{LK1} \right|^2 + \left| \alpha_j^{LK2} \right|^2 + 2\zeta \left| \alpha_j^{LK1} \right| \left| \alpha_j^{LK2} \right| \cos \left( \varphi_\mathbf{q} - \phi_{12}^j \right) \right\},
\end{aligned}
\qquad (11)
$$

where $\phi_{12}^j$ is the relative phase between $\alpha_j^{LK1}$ and $\alpha_j^{LK2}$.

Taking the low-energy limit of the effective TB model in ref. 1 (see equations (3)–(7)), one can derive the four effective velocities in the generalized Weyl Hamiltonian (equation (8)), which are given by

$$
\begin{aligned}
\mathbf{w}_0^{TB} &= (-5.06, \ 0.750), \\
\mathbf{v}^{TB} &= (6.70, \ 6.86) \quad (\text{in } 10^4 \text{ m s}^{-1}).
\end{aligned}
\qquad (12)
$$

Using these velocities, the phase $\varphi_\mathbf{q}$ in equation (11) can be approximated as $\varphi_\mathbf{q} \approx \arctan(q_y/q_x) (\equiv \varphi)$, where $\varphi$ is the angle between $\mathbf{q}$ and the $k_x$-axis. It is shown from equation (11) that the site-spectral weight $n_j^\zeta(\mathbf{q})$ acquires an anti-phase relation between the $j = B$ and the other sites, namely, $\phi_{12}^{A(A')} = \phi_{12}^C = 0$ and $\phi_{12}^B = \pi$. Moreover, $|\alpha_B^j| \sim 0.8$ and $|\alpha_C^j| \sim 0.7$ have an equal size for the bases $\lambda = \text{LK1}$ and $\lambda = \text{LK2}$, while one obtains $\left| \alpha_{A(A')}^{LK2} \right| \gg \left| \alpha_{A(A')}^{LK1} \right| \sim 0$ [or $\left| \alpha_{A(A')}^{LK1} \right| \gg \left| \alpha_{A(A')}^{LK2} \right| \sim 0$]. This causes an oscillation of equation (11) as a function of $\varphi$ around the DP, with a large amplitude and an opposite phase on the $j = B$-site and the $j = C$-site, whereas the $\varphi$ dependence is small on the $j = A(A')$ site (Supplementary Fig. 1a). As the cone is tilted in the $k_x$-direction (inset of Supplementary Fig. 1a), the Fermi velocity becomes highly anisotropic around the cone[20,25,50] (Supplementary Fig. 1a), and

can be expressed as

$$v_{\mathrm{F}}(\varphi,\zeta) = w_{0x}\cos\varphi + w_{0y}\sin\varphi + \zeta\sqrt{(v_x)^2\cos^2\varphi + (v_x)^2\sin^2\varphi}. \quad (13)$$

The most striking consequence of this anisotropy, in conjunction with equation (11), is that there is a large asymmetry in the site-spectral weight for the site $A(A')$, $B$ and $C$ around the DP. Namely, the site $B$ predominantly reflects the large-$v_F$ electrons in the steep slope of the cone ($\varphi \approx \pi$); the site $C$ mostly probes the small-$v_F$ electrons in the gentle slope ($\varphi \approx 0$); and the site $A(A')$ probes the entire electronic states on average around the DP (see Fig. 1d–f and Supplementary Fig. 1a,c). This asymmetry of the site-spectral weight results in a clear difference in the size of the $j$-site DOS, $D_j(E,\zeta)$, for the different sites—$D_C(E,\zeta) > D_{A(A')}(E,\zeta) > D_B(E,\zeta)$ (ref. 25)—where $D_j(E,\zeta)$ (per valley and ET molecule) is defined as the $\mathbf{q}$ summation of $n_j^\zeta(\mathbf{q})$ at a given energy in the band $\zeta$, which is expressed as

$$D_j(E,\zeta) = 2V_C \int \frac{d^2\mathbf{q}}{(2\pi)^2} n_j^\zeta(\varphi) \delta(E - E_\zeta(\mathbf{q})). \quad (14)$$

Here, $V_C$ is the 2D unit-cell volume in the conducting $ab$-plane. The contrasting features of the $B$ and $C$ site-spectral weights around the DP provide unique opportunities to probe the excitations of large-$v_F$ Dirac electrons (in the steep slope) and small-$v_F$ Dirac electrons (in the gentle slope) separately in terms of a site-selective local measurement such as NMR.

We note that in the original nonlinear TB calculation by Katayama et al.[25], the velocities are given in the unit of energy (meV), $\mathbf{w}_0 = (w_{0x},w_{0y}) = (-38.9, 4.8)$ and $\mathbf{v} = (v_x, v_y) = (51.5, 43.9)$, because both the primitive vectors and the reciprocal lattice vectors are set to have a unit length in their notation. To recover the ordinary physical unit (length/time), one has to multiply the velocities either by $a/\hbar$ or $b/\hbar$, using the values of the lattice constants $a$ and $b$ at the current pressure (2.3 GPa), where $\hbar$ is the Planck constant divided by $2\pi$. By linearly extrapolating the X-ray diffraction data obtained at 1.76 GPa (ref. 33) to 2.3 GPa, the lattice constants are estimated as $a = 8.567$ Å and $b = 10.282$ Å. The velocities in equation (12) are obtained in terms of these lattice constants.

**Renormalization-group calculations.** The observed nonlinear temperature dependence of the spin susceptibility below the inflection point $T'_{\mathrm{flex}}$ in Fig. 2d,e, cannot be understood within the non-interacting Dirac-fermion picture, as we mentioned above (see Methods: effective TB model). In this temperature range, the screening effect should be weak reflecting the vanishing thermal excitations around the gapless point at $E_F$. For this kind of situation, it is well known from the RG study of graphene Dirac cone that the LR part of the unscreened Coulomb interaction among electrons causes a logarithmic divergence of the Fermi velocity $v_F$ around the DP[7–14]. A similar argument has been recently proposed for the tilted Dirac cone in $\alpha$-I$_3$ (ref. 46), in which a RG flow of the Fermi velocity is suggested as a function of $T$. Hence, the most straightforward and reasonable way to understand the low-$T$ nonlinear feature of Fig. 2d would be to attribute it the $T$-driven renormalization of $v_F$ due to the LR Coulomb interaction.

To check this hypothesis, we have performed a RG calculation based on the generalized Weyl Hamiltonian (equation (8)) and tried to fit the data. A circular momentum cutoff of $\Lambda = 0.667$ Å$^{-1}$ around the DP ($\mathbf{q} = \mathbf{0}$) is introduced in the RG theory, which is of the size of the averaged inverse lattice constant, $\Lambda = 2\pi/L$ with $L = (a + b)/2 = 9.425$ Å at 2.3 GPa. For the initial values of the velocities at the cutoff momentum $q = \Lambda$ (that is, $\mathbf{v}$ and $\mathbf{w}_0$ in equation (8)), we employ velocities derived from the effective TB model of ref. 25, $\mathbf{w}_0^{\mathrm{TB}}$ and $\mathbf{v}^{\mathrm{TB}}$ (equation (12)), as discussed in the previous subsection (Methods: Generalized Weyl Hamiltonian and site-spectral weight). In the one-loop order large-$N$ expansion of the RG theory, $\mathbf{v} = (v_x, v_y)$ are renormalized following equation (2) (given in the main text) and grows logarithmically as functions of $\Lambda/q$ (where $q$ is measured from the DP), whereas $\mathbf{w}_0 = (w_{0x}, w_{0y})$ are not renormalized (Supplementary Fig. 4a). We note that equation (2) takes into account the screening effect of the Coulomb interaction including the polarization bubbles in the self-energy. It is applicable to any size of the coupling $g_0 = 2\pi\varepsilon^2 N / (16\varepsilon\sqrt{v_x^2\cos^2\varphi + v_y^2\cos^2\varphi})$, where $\varepsilon$ is the dielectric constant and $N = 4$ is the number of fermion species corresponding to the two DPs in the Brillouin zone and two spin projections (that means the twofold valley degeneracy is considered).

Reflecting the renormalization of $\mathbf{v}$, the eigenenergy with the RG correction becomes (with the band index $\zeta = \pm$)

$$\widehat{E_\zeta(\mathbf{q})} = \hbar\left(\mathbf{w}_0 \cdot \mathbf{q} + \zeta\sqrt{v_x(q)^2 q_x^2 + v_y(q)^2 q_y^2}\right). \quad (15)$$

Correspondingly, the local DOS and the electron-spin susceptibility for the site $j$ are respectively given by the expressions

$$\widehat{D_j(E,\zeta)} = 2V_C \int \frac{d^2\mathbf{q}}{(2\pi)^2} n_j^\zeta(\varphi) \delta\left(E - \widehat{E_\zeta(\mathbf{q})}\right), \quad (16)$$

$$\widehat{\chi_s^j(T)} = \mu_B \frac{\partial}{\partial H} \sum_{\zeta=\pm} \frac{N}{2} V_C \int_{-\infty}^{\infty} dE \int \frac{d^2\mathbf{q}}{(2\pi)^2} n_j^\zeta(\varphi) \delta\left(E - \widehat{E_\zeta(\mathbf{q})}\right)$$
$$\left\{ f\left(E - \frac{E_Z}{2}\right) - f\left(E + \frac{E_Z}{2}\right) \right\}, \quad (17)$$

where $V_C = 88.086$ Å$^2$ is the 2D unit-cell volume in the conducting $ab$-plane estimated at 2.3 GPa, $N = 4$ is the number of fermion species (again, the twofold valley degeneracy is included), $E_Z = g\mu_B H$ is the electron Zeeman energy and $f(E)$ is the Fermi distribution. The integration with respect to $\mathbf{q}$ in equations (16) and (17) is done up to the momentum cutoff ($q = \Lambda$). Note that $T$ plays the role of the flow parameter in equation (17) (that is, a $T$-driven RG flow).

In Supplementary Figs 4–6, we present the calculated profiles of the velocities (Supplementary Fig. 4), the $j$-site DOS (Supplementary Fig. 5) and the $j$-site electron-spin susceptibility (Supplementary Fig. 6) based on the RG equation (equation (2)) and equations (15)–(17). Here, to get a reasonable agreement with the experiment, we introduced a phenomenological parameter, $u$ ($\leq 1$), in the calculations which is defined by the expressions

$$\mathbf{w}_0' = u\mathbf{w}_0^{\mathrm{TB}} \quad \text{and} \quad \mathbf{v}' = u\mathbf{v}^{\mathrm{TB}}. \quad (18)$$

This parameter reflects a suppression of the velocity or a reduction of the hopping amplitude $t_{ij}$ between the lattice site $i$ and $j$ due to the SR part of the Coulomb interaction[53], as mentioned in the main text. Then, the RG flow, which is determined by equation (2), is controlled by two parameters—the dielectric constant $\varepsilon$ and the phenomenological parameter $u$.

Supplementary Fig. 4b,c, presents the parameter dependence of the flow of the velocities $v_x$ and $v_y$. The dielectric constant $\varepsilon$ affects the power of the flow function $\mathbf{v} = \mathbf{v}(\Lambda/q)$ (Supplementary Fig. 4b), while the parameter $u$ determines both the initial values of the velocities and the power of the flow (Supplementary Fig. 4c). Supplementary Fig. 5b,c, depicts the $\varepsilon$ and $u$ dependence of the DOS $D_j(E,\zeta)$ for the site $j = A(A')$. It is shown that a prominent suppression of the DOS develops around $E = E_F$ ($= 0$) by reducing the dielectric constant $\varepsilon$ (which corresponds to an increase in the coupling $g_0$ at the cutoff momentum $q = \Lambda$). The reduction of the parameter $u$, on the other hand, leads to an enhancement of the DOS because the DOS is linked to the inverse square of the velocities[20]. In Supplementary Fig. 6b,c, we show the parameter dependence of the calculated electron-spin susceptibility. As the susceptibility probes the $k_BT$-average of the DOS near $E_F$ ($= 0$), the suppression of the DOS for small $\varepsilon$ causes a reduction of the susceptibility at low temperatures (Supplementary Fig. 6b). The enhancement of the susceptibility for a small value of $u$ can be also understood in the same fashion (Supplementary Fig. 6c).

Supplementary Fig. 7a shows the result of the variance analyses of the $T$-driven RG-flow fit to the experimental spin susceptibilities for the site $j = A(A')$ and $C$ (Fig. 2d) based on equation (17). Here, the variance, plotted in the $u - \varepsilon - \mathrm{Var}(\chi_s^j)$ space, is defined by

$$\mathrm{Var}(\chi_s) = \sum_{j=A(A'),C} \frac{1}{n} \sum_{i=1}^{n} \left( \chi_s^j(T_i) - \widehat{\chi_s^j(T_i)} \right)^2, \quad (19)$$

where $T_i$ are the experimental temperature points ($i = 1,2, \cdots, n$), and $\chi_s^j(T_i)$ and $\widehat{\chi_s^j(T_i)}$ are the measured and calculated susceptibilities at $T = T_i$, respectively. Note that the variance is defined only for the results on the site $A(A')$ and $C$, since the results on the site $B$ has less good agreement with the RG-calculation (see Supplementary Fig. 7b–e). This is because of the emergent ferrimagnetic spin polarization—the negative susceptibility on the site $B$ at low temperatures (inset of Fig. 3c)—which is due to the SR part of the Coulomb interaction, not considered in this calculation. (For details, see Methods: Simulations with the Hubbard model.) The calculated susceptibilities at selected values of $u$ and $\varepsilon$ are shown for three different sites $j = A(A')$, $B$ and $C$ together with the experimental data in Supplementary Fig. 7b–e. Except for the site $B$, the agreement with the calculations and the experiments is pretty good for small values of $u$ and $\varepsilon$. The variance $\mathrm{Var}(\chi_s^j)$ has a minimum around $(u, \varepsilon) = (0.35, 1)$ with little $\varepsilon$ dependence up to $\varepsilon \approx 10^{1.5}$ (Supplementary Fig. 7b,c) and increases rapidly as one moves away from this minimum especially when $u$ is increased. Thus, we take these values of $(u, \varepsilon) = (0.35, 1)$ as the optimal parameters hereafter. (Here, the result for $\varepsilon = 1$ is chosen because the $\varepsilon$-dependence is small and affects the result little.) The small value of $u = 0.35(< 1)$ is in agreement with the aforementioned reduction of the hopping amplitude due to the SR repulsion[53] and indicates the presence of moderate electron correlations. (This point naturally supports our discussion based on the Hubbard model in the next subsection to deal with the observed negative spin susceptibility on the site $B$.)

Lastly, it may be worthwhile to mention that the flow of the velocities $\mathbf{v} = (v_x, v_y)$ with $\mathbf{w}_0 = (w_{0x}, w_{0y})$ staying constant in Supplementary Fig. 4a leads to a situation where $v_x, v_y \gg |w_{0x}|, w_{0y}$ is realized for a large value of the momentum scale $\Lambda/q$ ($\gg 1$) (that is, in the vicinity of the DP). This means that the tilting term ($\mathbf{w}_0$) becomes effectively negligible with respect to the anisotropy term ($\mathbf{v}$) at low energy in the generalized Weyl Hamiltonian (equation (8)). Moreover, the values of the two velocities, $v_x$ and $v_y$, remain very close to each other down to the vicinity of the DP (for instance, $\ln(\Lambda/q) = 5$ in Supplementary Fig. 4a corresponds to $q = 0.0045$ Å$^{-1}$ in terms of $\Lambda = 0.667$ Å$^{-1}$). These points together suggest that the anisotropy of the Dirac cone becomes very small near the DP and the cone is practically isotropic at low energy, as one can see in the calculated cone in Fig. 4d–f. This is can will account for the observed site dependence of the susceptibility, the root of which is linked to the tilt and the anisotropy of the cone. The site dependence becomes vanishingly small at low temperatures (Supplementary Fig. 2), which reflect the renormalization of the velocity that makes the cone to be more and more isotropic at low energy.

**Simulations with the Hubbard model.** In the previous subsection, we have shown that the RG calculation based on the low-energy continuum model (equation (8)) well reproduces the observed nonlinear $T$-dependence of $\chi_s^j(T)$ at the site $A(A')$ and $C$ (Supplementary Fig. 7b–e) when the velocity suppression due to the SR Coulomb interaction is phenomenologically taken into account. However, the agreement is less good at the site $B$; in particular, the negative susceptibility $\chi_s^B < 0$ below $T \approx 60$ K cannot be reproduced at all, suggesting the presence of other interaction effects.

To understand the origin of the observed $\chi_s^B < 0$, we have performed another simulation of the susceptibility based on the RPA. For this, we start from the standard Hubbard model with the onsite Hubbard interaction $U$, by extending earlier study[50]. The model Hamiltonian is given for the present system by the expression

$$H = \sum_{(i\alpha;j\beta),\sigma} \left( t_{i\alpha;j\beta} a_{i\alpha\sigma}^{\dagger} a_{j\beta\sigma} + \text{h.c.} \right) + \sum_{j\alpha} U a_{j\alpha\uparrow}^{\dagger} a_{j\alpha\uparrow} a_{j\alpha\downarrow}^{\dagger} a_{j\alpha\downarrow}, \quad (20)$$

where $a_{j\alpha\sigma}^{\dagger}$ is the creation operator on the site $j = A(1)$, $A'(2)$, $B(3)$ and $C(4)$ in the unit cell $\alpha (= 1, \ldots, N_{\text{u.c.}})$ with the spin $\sigma (= \uparrow, \downarrow)$, $N_{\text{u.c.}}$ is the total number of the unit cell, and $t_{i\alpha;j\beta}$ is the nearest-neighbour and next nearest-neighbour hopping amplitude between the lattice site $(i, \alpha)$ and $(j, \beta)$ (see Supplementary Fig. 8). (It should be noticed that this Hamiltonian (equation (20)) lacks the site-dependent potential term (equations (3) and (4)) we employed above. From qualitative points of view, the inclusion of these potentials does not alter the results much, and, for the sake of simplicity, we shall omit this term in this subsection.)

The amplitude $t_{i\alpha;j\beta}$ at finite temperatures are estimated from the hopping integrals given by the first-principle calculations[24] at 8K, $\{t_p^{\text{LT}}; p = \text{a1} - \text{a4}\}$ (equations (6) and (7)), and at 300 K, $\{t_p^{\text{HT}}; p = \text{a1} - \text{a4}\}$, as given in the following (in eV)

$$
\begin{aligned}
t_{\text{a1}}^{\text{HT}} &= -0.0101, & t_{\text{a2}}^{\text{HT}} &= -0.0476, & t_{\text{a3}}^{\text{HT}} &= 0.0093, \\
t_{\text{b1}}^{\text{HT}} &= 0.1081, & t_{\text{b2}}^{\text{HT}} &= 0.1109, & t_{\text{b3}}^{\text{HT}} &= 0.0551, & t_{\text{b4}}^{\text{HT}} &= 0.0151, \\
t_{\text{a1'}}^{\text{HT}} &= 0.0088, & t_{\text{a3'}}^{\text{HT}} &= 0.0019, & t_{\text{a4'}}^{\text{HT}} &= 0.0009,
\end{aligned} \quad (21)
$$

in combination with the interpolation formula given in ref. 50

$$t_p(T) = t_p^{\text{LT}} + \left( t_p^{\text{HT}} - t_p^{\text{LT}} \right) \frac{T - 8}{300 - 8}. \quad (22)$$

Within the mean-field approximation, the diagonalization of the Hubbard Hamiltonian (equation (20)) yields

$$\sum_{j=1}^{4} \tilde{\epsilon}_{ij\sigma}(\mathbf{k}) d_{j\eta\sigma}(\mathbf{k}) = E_{\eta\sigma}(\mathbf{k}) d_{i\eta\sigma}(\mathbf{k}), \quad (23)$$

$$\tilde{\epsilon}_{ij\sigma}(\mathbf{k}) = \epsilon_{ij}(\mathbf{k}) + U\langle N_{i\sigma}\rangle\delta_{ij}, \quad (24)$$

$$\langle N_{j\sigma}\rangle = \frac{1}{N_{\text{u.c.}}} \sum_{\mathbf{k}} \sum_{n=1}^{4} d_{jn,-\sigma}^*(\mathbf{k}) d_{jn,-\sigma}(\mathbf{k}) f\left( E_{n,-\sigma}(\mathbf{k}) - \mu \right), \quad (25)$$

where we define $\epsilon_{ij}(\mathbf{k}) = \sum_{\delta_{ij}} t_{ij} e^{\mathbf{k}\cdot\delta_{ij}}$, $\boldsymbol{\delta}_{ij}$ is a vector connecting the nearest-neighbour lattice sites $i$ and $j$, $E_{n\sigma}(\mathbf{k})$ is the eigenvalue ($E_{1\sigma} > E_{2\sigma} > E_{3\sigma} > E_{4\sigma}$), $d_{i n\sigma}(\mathbf{k})$ is the corresponding eigenvector, $f(E)$ is the Fermi distribution and $\mu$ is the chemical potential. Note that the average electron number $\langle N_{j\sigma}\rangle$ is determined self-consistently from the condition $\sum_{j\sigma} \langle N_{j\sigma}\rangle = 6$, reflecting the $\frac{3}{4}$-filling of the band. In the normal state, one has $\langle N_{j\uparrow}\rangle = \langle N_{j\downarrow}\rangle$; thus, the spin $\sigma$ is omitted hereafter.

We introduce the bare site-spin susceptibility matrix, $\hat{\chi}^{(0)}$, whose $(ij)$-element is given by

$$\chi_{ij}^{(0)}(\mathbf{Q}, \omega) = -\frac{1}{N_{\text{u.c.}}} \sum_{\mathbf{k}} \sum_{\eta,\eta'=1}^{4} \mathcal{F}_{ij}^{\eta\eta'}(\mathbf{k}, \mathbf{Q}) \frac{f(E_\eta(\mathbf{k} + \mathbf{Q})) - f(E_{\eta'}(\mathbf{k}))}{E_\eta(\mathbf{k} + \mathbf{Q}) - E_{\eta'}(\mathbf{k}) - \hbar\omega - i\delta}, \quad (26)$$

in terms of a form factor

$$\mathcal{F}_{ij}^{\eta\eta'}(\mathbf{k}, \mathbf{Q}) = d_{i\eta}(\mathbf{k} + \mathbf{Q}) d_{j\eta}^*(\mathbf{k} + \mathbf{Q}) d_{j\eta'}(\mathbf{k}) d_{i\eta'}^*(\mathbf{k}), \quad (27)$$

where $i\delta$ ($\delta > 0$) is an infinitesimally small imaginary part. Within the RPA approach, the spin fluctuations are estimated using the expression[50]

$$\chi_{ij,\text{RPA}}(\mathbf{Q}, \omega) = (\hat{\chi}_{\text{RPA}})_{ij}(\mathbf{Q}, \omega) = \left[ \left( \hat{I} - \hat{\chi}^{(0)} U\hat{I} \right)^{-1} \hat{\chi}^{(0)} \right]_{ij}(\mathbf{Q}, \omega), \quad (28)$$

(where $\hat{I}$ is the $4 \times 4$ unit matrix), and the total RPA $j$-site-spin susceptibility (for $\mathbf{Q} = \mathbf{0}$ and $\omega = 0$) is given by

$$\chi_{\text{RPA}}^j = \sum_{i=1}^{4} \left[ \left( \hat{I} - \hat{\chi}^{(0)} U\hat{I} \right)^{-1} \hat{\chi}^{(0)} \right]_{ji}(\mathbf{0}, 0). \quad (29)$$

Now, we decompose the bare susceptibility $\chi_{ij}^{(0)}$ into two parts (for the reason that will become clear below): $\chi_{ij}^{(0)} = \chi_{ij}^{(0),\text{intra}} + \chi_{ij}^{(0),\text{inter}}$, where $\chi_{ij}^{(0),\text{intra}}$ and $\chi_{ij}^{(0),\text{inter}}$ correspond to the intra-band and inter-band contributions to the bare susceptibility (for $\mathbf{Q} = \mathbf{0}$), respectively. The intra-band RPA susceptibility is then defined by

$$\chi_{\text{RPA}}^{j,\text{intra}} = \sum_{i=1}^{4} \left[ \left( \hat{I} - \hat{\chi}^{(0),\text{intra}} U\hat{I} \right)^{-1} \hat{\chi}^{(0),\text{intra}} \right]_{ji}(\mathbf{0}, 0), \quad (30)$$

and the inter-band contribution to the total RPA susceptibility is expressed as

$$\chi_{\text{RPA}}^{\text{inter}} = \chi_{\text{RPA}}^j - \chi_{\text{RPA}}^{j,\text{intra}}. \quad (31)$$

In Supplementary Fig. 3a, the calculated temperature dependence of the intra-band RPA susceptibility $\chi_{\text{RPA}}^{j,\text{intra}}$ is shown for $U = 0.14$ eV in comparison to the non-interacting case ($U = 0$) for the site $j = A(A')$, $B$ and $C$. It is seen that the intra-band susceptibility $\chi_{\text{RPA}}^{j,\text{intra}}$ for a finite value of $U$ becomes always larger than the case for $U = 0$ at all temperature and all site $j$. The inter-band contribution, on the other hand, is found to provide a site-dependent correction to the susceptibility (Supplementary Fig. 3b). Namely, the inter-band RPA susceptibility $\chi_{\text{RPA}}^{j,\text{inter}}$ gives a positive contribution on the site $A(A')$ and $C$, whereas the contribution is negative on the site $B$. The negative inter-band contribution on the site $B$ ($\chi_{\text{RPA}}^{B,\text{inter}} < 0$) develops with increasing $U$ and in turn leads to a negative total RPA susceptibility ($\chi_{\text{RPA}}^B < 0$) above a critical $U$ value $U_C \approx 0.12$ eV. The position of the minimum shifts towards higher energies with increasing $U$ (Supplementary Fig. 3c). Thus, by taking a value of $U = 0.14$ eV, the minimum of the total RPA susceptibility $\chi_{\text{RPA}}^B$ appears at around $T \approx 50$ K, which agrees well with the experiment (inset of Fig. 3c). These results demonstrate that the SR part of the Coulomb interaction between electrons causes a ferrimagnetic spin polarization at low temperature. This leads to a situation where the site $B$ is subjected to a negative local magnetic field, pointing opposite to the external field, while the other sites ($A$, $A'$ and $C$) feel a positive field, as schematically illustrated in Fig. 6.

To have an overall comparison of the RPA calculations with the experiment, the first derivative of the susceptibility is analysed for the case $U = 0$, the total RPA susceptibility $\chi_{\text{RPA}}^j$ (equation (29)) for $U = 0.14$ eV, and the observed spin susceptibility (Fig. 2e), as depicted for the non-equivalent sites in Supplementary Fig. 3d–f. The calculations capture the experimental features relatively well on the site $A(A')$ and $C$ above the peak temperature ($T \approx 50$ K), whereas the calculation does not agree with the experiment at all on the site $B$. At low temperatures, on the other hand, the agreement between the calculation and experiment becomes worse even for the site $A(A')$ and $C$. That is, in the low temperature limit, the calculations show a saturation both for $U = 0$ and the finite $U$, while the experiment exhibits a monotonic decrease towards zero (Supplementary Fig. 3d and f). We believe that this disagreement arises because the present RPA calculation does not incorporate the $T$-driven $v_F$-renormalization effect due to the LR part of the Coulomb interaction, as discussed in the previous subsection (see Methods: RG calculations). The $v_F$-renormalization results in a super-linear temperature dependence in the spin susceptibility $\chi_s^j$ (Supplementary Fig. 6), which causes a decrease of the first derivative of $\chi_s^j$ with decreasing temperature as reflected in the experiment.

Finally, we comment on the comparison of the experiment with an orthodox RPA fitting. This is done by assuming a simplified RPA (s-RPA) expression for the spin susceptibility, which is defined by

$$\chi_{\text{sRPA}}^j = \frac{\chi_j^{(0)}}{1 - U_j\chi_j^{(0)}}, \quad (j = A, A', B \text{ and } C) \quad (32)$$

where $\chi_j^{(0)}$ is the bare spin susceptibility and $U_j$ is an adjustable parameter reflecting the onsite Hubbard interaction. (Note that $\chi_j^{(0)}$ and $\chi_{\text{sRPA}}^j$ correspond to the diagonal term in equations (26) and (28), respectively, for $\mathbf{Q} = \mathbf{0}$ and $\omega = 0$ with the Hubbard interaction $U$ in equation (20) replaced by the site-dependent parameter $U_j$.) Supplementary Fig. 14 presents the least-square fitting results to the experiment using the s-RPA expression, which yields $U_{A(A')} = 0.6$, $U_B = 1.3$ and $U_C = 0.4$ (in eV). It is clearly seen that the observed nonlinear temperature dependence of the susceptibility cannot be reproduced by the s-RPA fit at all sites. In particular, there is an unphysical divergence in the calculation, which is linked to the large $U_j$ values used in the calculation that are too large compared with the typical values applicable to this system[24,26,31,35,40,41,48,50]. It is thus concluded that one has to consider the full-matrix elements of equation (28) in order to obtain a reasonable agreement with the experiment.

**Data availability.** The data that support the findings of this study are available from M.H. upon requests.

## Acknowledgements

We gratefully acknowledge valuable discussions with Y. Suzumura, N. Nagaosa, H. Isobe, M. Imada, M. Ogata, H. Matsuura, H. Fukuyama, C. Hotta, S. Sugawara, T. Osada, T. Taniguchi, M.-H. Julien and H. Mayaffre. In particular, we thank M. Horvatić and M. O. Goerbig for their thoughtful advices on the analyses and dedicated discussions. We also thank D. Liu for providing us unpublished results and for fruitful discussions. This work was supported by MEXT/JSPJ KAKENHI under Grant Noes 20110002, 21110519, 24654101, 25220709, 15K05166, 15H02108, JSPS Postdoctoral Fellowship for Research Abroad (Grant No. 66, 2013) and MEXT Global COE Program at University of Tokyo (Global Center of Excellence for the Physical Sciences Frontier; Grant No. G04).

## Author contributions

The samples were prepared by M.T. The data were taken, analysed and interpreted by M.H. and K.I. with the help of K.M., C.B., D.B., G.M., A.K. and K.K. The renormalization-group calculation was done by D.B. and analysed by M.H. with the help of D.B. and C.B. The simulations using the Hubbard model were carried out by G.M. and A.K. The project was supervised by K.K, and the manuscript was written by M.H. with K.M., D.B., A.K., C.B. and K.K.

## Additional information

**Supplementary Information** accompanies this paper at http://www.nature.com/naturecommunications

**Competing financial interests:** The authors declare no competing financial interests.

**Reprints and permission** information is available online at http://npg.nature.com/reprintsandpermissions/

**How to cite this article:** Hirata, M. *et al.* Observation of an anisotropic Dirac cone reshaping and ferrimagnetic spin polarization in an organic conductor. *Nat. Commun.* 7:12666 doi: 10.1038/ncomms12666 (2016).